\documentclass[floatfix,aps,prb,twocolumn, showpacs]{revtex4-1}

\makeatletter
\newsavebox{\@brx}
\newcommand{\llangle}[1][]{\savebox{\@brx}{\(\m@th{#1\langle}\)}%
  \mathopen{\copy\@brx\kern-0.5\wd\@brx\usebox{\@brx}}}
\newcommand{\rrangle}[1][]{\savebox{\@brx}{\(\m@th{#1\rangle}\)}%
  \mathclose{\copy\@brx\kern-0.5\wd\@brx\usebox{\@brx}}}
\makeatother

\usepackage{hyperref}
\usepackage{amssymb}
\usepackage{amsmath}
\usepackage{verbatim}
\usepackage[dvips]{graphicx}
\usepackage[utf8]{inputenc}
\usepackage{bm}
\usepackage{float}
\usepackage{color}

\newcommand{\be}{\begin{equation}}
\newcommand{\ee}{\end{equation}}
\newcommand{\bea}{\begin{eqnarray}}
\newcommand{\eea}{\end{eqnarray}}

\begin{document}

\title{Anisotropy and spin-fluctuation effects on the spectral properties of Shiba impurities}

\author{J. A. Andrade}
\affiliation{Facultad de Ciencias Exactas y Naturales, Universidad Nacional de Cuyo, 5500  Mendoza, Argentina}
\affiliation{Instituto Interdisciplinario de Ciencias B\'asicas, Consejo Nacional de Investigaciones Cient\'{\i}ficas y T\'ecnicas (CONICET), 5500 Mendoza, Argentina}
\author{Alejandro M. Lobos}
\email{alejandro.martin.lobos@gmail.com}
\affiliation{Facultad de Ciencias Exactas y Naturales Universidad Nacional de Cuyo, 5500  Mendoza, Argentina}
\affiliation{Instituto Interdisciplinario de Ciencias B\'asicas, Consejo Nacional de Investigaciones Cient\'{\i}ficas y T\'ecnicas (CONICET), 5500 Mendoza, Argentina}

\begin{abstract}
We theoretically consider a quantum magnetic impurity coupled to a superconductor, and obtain the local density of states at the position of the impurity taking into account the effect of spin-fluctuations  and  single-ion magnetic anisotropy. We particularly focus on the spectrum of subgap Yu-Shiba-Rusinov (YSR or Shiba) states induced by a quantum impurity with easy- or hard-axis uniaxial anisotropy. Although this is a relevant experimental situation in, e.g.,  magnetic adatoms on the surface of clean metals, it is customary that theoretical descriptions assume a classical-spin approximation which is not able to account for single-ion anisotropy and other quantum effects. Here, quantum fluctuations of the spin are taken into account in the equations of motion of the electronic Green's function in the weak-coupling limit, and considerably modify the energy of the Shiba states compared to the classical-spin approximation. Our results point towards the importance of incorporating quantum fluctuations and anisotropy effects for the correct interpretation of scanning tunneling microscopy (STM) experiments.
\end{abstract}


\maketitle

\section{Introduction}\label{intro}

The competition between superconductivity and magnetism at the atomic scale gives rise to interesting and exotic quantum phenomena\cite{Balatsky_2006,Heinrich18_Review_single_adsorbates}. Magnetic impurities adsorbed on the surface of clean superconductors are a physical realization where this competition can be studied experimentally using, for instance, low-temperature scanning-tunneling microscopy (STM) techniques. The local STM differential conductance $dI/dV$ near the impurity reveals the presence of the so-called Yu-Shiba-Rusinov, or simply Shiba, states which emerge due to the disruption of the superconducting state produced by the local exchange field of the impurity. Originally predicted in seminal papers\cite{Yu65_YSR_states, Shiba_1968, Rusinov69_YSR_states}, Shiba states appear as resonances in the STM differential conductance, symmetrically located around the Fermi level at energies within the superconducting gap $\Delta$, and localize around the impurity\cite{Yazdani97_YSR_states}. Recent progress in STM techniques has shown  a surprisingly complex behavior of Shiba states, as a result of the interplay between quantum fluctuations, Kondo screening, single-ion anisotropy, etc.\cite{Ji08_YSR_states, Iavarone10_Local_effects_of_magnetic_impurities_on_SCs, Ji10_YSR_states_for_the_chemical_identification_of_adatoms, Franke_2011, Bauer13_Kondo_screening_and_pairing_on_Mn_phtalocyanines_on_Pb, Hatter_2015,Hatter2017_Scaling_of_YSR_energies, Ruby_2016, Choi_2017,
Ruby15_Tunneling_into_localized_subgap_states_in_SC}

Due to their fascinating properties, Shiba states have been the focus of a growing number of experimental and theoretical works. Since they are naturally protected from decoherence by the presence of the superconducting gap
\cite{Heinrich13_Protection_of_excited_spin_states_by_SC_gap}, they have become increasingly attractive from the point of view of novel quantum information and quantum processing technologies. In addition, recent theoretical proposals made the striking prediction that  hybridization of Shiba states can lead to ``Shiba bands'' with nontrivial topological character and to the emergence of Majorana zero-modes in one-dimensional chains of magnetic atoms\cite{Nadj-Perdge13_Majorana_fermions_in_Shiba_chains, Klinovaja13_TSC_and_Majorana_Fermions_in_RKKY_Systems, Braunecker13_Shiba_chain, Pientka13_Shiba_chain}. Subsequent STM experiments realized on Fe atomic chains deposited ontop of superconducting Pb(111) or Pb(110) surfaces have revealed intriguing zero-bias peaks in the $dI/dV$ signal, consistent with the Majorana zero-mode scenario   \cite{NadjPerge14_Observation_of_Majorana_fermions_in_Fe_chains, Pawlak15_Probing_Majorana_wavefunctions_in_Fe_chains, Ruby_2015}.

When considering a magnetic impurity in a superconductor, it is customary to make the simplifying assumption that the impurity spin $S$ is a classical object (essentially, a point-like magnetic field with no internal dynamics), a situation which  is physically expected in the large-spin limit $S\rightarrow \infty$. However, it is well-known that the classical-spin approximation cannot describe the experimentally observed Kondo effect \cite{Franke_2011, Bauer13_Kondo_screening_and_pairing_on_Mn_phtalocyanines_on_Pb, Hatter_2015, Hatter2017_Scaling_of_YSR_energies}, an inherently  quantum many-body phenomenon\cite{Kondo, Hewson_1993}. Phenomenologically, in the case of an isotropic spin $S$ coupled to a single-band superconductor, the  ground state of the full system (superconductor plus impurity)  depends on the competition between the Kondo effect and pairing correlations\cite{Zittartz_1970_I, Zittartz_1970_III, Yoshioka98_Kondo_impurity_in_SC_with_NRG,Yoshioka00_NRG_Anderson_impurity_on_SC,Franke_2011}. When the Kondo temperature $T_K\gg \Delta$, the many-body ground state is a Kondo-screened state 
with total spin $S_\text{T}=S-1/2$ and odd-fermion parity. The first excited  (many-body) state corresponds to an unscreened $S_\text{T}=S$  multiplet with even fermion parity. This situation is reversed when $\Delta \gg T_K$. 
A quantum phase transition (QPT) between these two ground states occurs at $T_K\sim 0.3 \Delta$\cite{Satori92_Magnetic_impurities_in_SC_with_NRG, Sakai93_Magnetic_impurities_in_SC_with_NRG,Yoshioka98_Kondo_impurity_in_SC_with_NRG,Yoshioka00_NRG_Anderson_impurity_on_SC}, and is signalled  by the crossing of the Shiba states at the Fermi energy, a feature that allows its experimental detection by STM techniques\cite{Franke_2011,Hatter_2015}. This transition is also known as the ``$0-\pi$" transition in the context of electronic transport through quantum dots attached to superconducting leads \cite{MartinRodero11_Review_Josephson_and_Andreev_transport_through_QDs}.

Single-ion magnetic anisotropy is another effect that profoundly modifies  the behavior of quantum impurities at low temperatures\cite{Zitko_2011, Zitko_2017}. It arises due to the presence of strong spin-orbit coupling and lack of inversion symmetry at the surface of clean metals, and therefore it is ubiquitous in magnetic-adatom systems studied with STM techniques\cite{Ji08_YSR_states, Iavarone10_Local_effects_of_magnetic_impurities_on_SCs, Ji10_YSR_states_for_the_chemical_identification_of_adatoms, Hatter_2015, Ruby_2016, Choi_2017}. 
These experimental systems show a complex subgap electronic structure with multiple Shiba states,  and demand for theoretical approaches that can go beyond the classical-spin approximation for their understanding. In that respect, it is interesting to mention recent works which consider the quantum nature of magnetic impurities, either by using the exact, but numerically costly, numerical renormalization group (NRG) method\cite{Bauer07_NRG_Anderson_model_in_BCS_superconductor, Zitko_2011, Zitko16_Spectral_properties_of_Shiba_states_at_finite_T, Zitko_2017}, or perturbation approaches in the parameter $U$ of the  Anderson model\cite{Zonda16_Perturbation_theory_for_Anderson_impurity_in_SC, Janis16_0_Pi_Transition_within_perturbation_theory, Zonda16_Perturbation_theory_of_SC_0_Pi_transition}, which limited their attention to the ideally isotropic case. The Anderson model in a superconducting host has also been studied by means of the  quantum Monte Carlo method\cite{Luitz10_QMC_study_of_Anderson_impurity_on_SC}, but its intrinsic difficulty to perform the analytical continuation to real frequencies prevents the use of this technique in this case, where the sharp Shiba resonances need to resolved. In fact, there are actually few theoretical methods that can reliably account for experimentally relevant effects, such as spin fluctuations, anisotropy and temperature. 

In this work we study a fully quantum spin $S$ coupled to a superconducting host via a $s$-$d$ exchange coupling term, and consider the effect of uniaxial anisotropy and finite temperature. We implement a novel decoupling scheme of the equations of motion for the electronic Green's function, formally valid in the unscreened regime $T_K\ll \Delta$ where the coupling to the superconductor is weak. Our results are consistent with previous works\cite{Zitko_2011, Zitko_2017} and point to the importance of quantum fluctuations and anisotropy for the low-temperature properties of Shiba impurities. 
For an impurity with easy-axis anisotropy, when either the spin $S\rightarrow\text{\ensuremath{\infty}}$
or the anisotropy parameter $D\rightarrow \infty$, the classical limit for the Shiba-state energy is recovered. However, for realistic values of $S$, the position of the Shiba state strongly depends on the value of $D$, and can differ considerably from the value predicted classically. Moreover, when $D<0$ (hard-axis anisotropy) the impurity \emph{never reaches the classical limit}, since the impurity spin becomes effectively $S_\text{eff}\rightarrow 0$ for integer spins  ($S_\text{eff}\rightarrow 1/2$ for half-integer spins) at low temperatures, contradicting the ``large'' spin hypothesis. Finally, at finite temperatures important deviations from the classical value are obtained, an effect that cannot be reproduced within the classical-spin approximation. 

The rest of the paper is organized as follows. In Sec. \ref{sec:model} we present the theoretical model and provide a short overview of previous theoretical results. In Sec. \ref{sec:decoupling_eom} we present our decoupling scheme for the Green's functions equations of motion and give details on its numerical resolution. In Sec. \ref{sec:results_T0} we show our results for the energy of the Shiba states obtained in different temperature and anisotropy regimes. Finally, in Sec. \ref{sec:summary} we give a  summary and some perspectives.

\section{Theoretical model and overview of previous results}\label{sec:model}

We theoretically describe a magnetic impurity deposited ontop a clean superconductor by the following Hamiltonian
\begin{align}
H &=H_{\text{SC}}+H_\text{s-d}+H_{\text{anis}}.\label{eq:H_total-1}
\end{align}
Here $H_\text{SC}$ is the BCS Hamiltonian describing a two-dimensional (2D) $s$-wave superconducting  film 
\begin{align}
H_{\text{SC}}&=\sum_{\mathbf{k}\sigma} \varepsilon_{\mathbf{k}}c_{\mathbf{k}\sigma}^{\dagger}c_{\mathbf{k}\sigma}+\Delta\sum_{\mathbf{k}}\left(c_{\mathbf{k}\uparrow}^{\dagger}c_{-\mathbf{k}\downarrow}^{\dagger}+c_{-\mathbf{k}\downarrow}c_{\mathbf{k}\uparrow}\right) \label{HSC},
\end{align}
where $c_{\mathbf{k},\sigma}^{\dagger}$($c_{\mathbf{k},\sigma}$) creates (annihilates)
an electron  in the conduction band  with 2D quasi-momentum $\mathbf{k}$ and spin $\sigma=\left\{\uparrow,\downarrow\right\}$ along the $\hat{z}$ axis (assumed perpendicular to the surface), $\varepsilon_\mathbf{k}$ is the dispersion relation of normal quasiparticles, and $\Delta$ is the superconductor pairing potential, which we take as the unit of energies in the rest of this work. 
The assumption of a 2D superconductor is not essential here, but it greatly simplifies the theoretical description since the translational symmetry along the $\hat{z}$-axis is broken by the surface, and the quasi-momentum $k_z$ becomes a non-conserved quantity. In addition, we assume a temperature-independent pairing parameter $\Delta$ [which can be taken from $\Delta\left(T\rightarrow 0\right)$ in STM experiments] in order to avoid solving the BCS gap-equation at finite temperatures \cite{Tinkham_Introduction_to_superconductivity}. This comes at the price of having to restrict the temperature to the regime $T\lesssim T_c/2$, where $T_c$ is the BCS critical temperature, where this approximation is well justified. As we will see later, this limitation is not serious.

The microscopic coupling of the magnetic impurity to the superconducting film is given  by the single channel anisotropic  $s$-$d$ exchange (or Kondo) Hamiltonian\cite{Balatsky_2006, Hewson_1993}
\begin{align}
H_\text{s-d} &=\frac{1}{V}\sum_{\mathbf{k},\mathbf{k}^{\prime}}\left[J_\parallel S_{z}\frac{c_{\mathbf{k}\uparrow}^{\dagger}c_{\mathbf{k}^{\prime}\uparrow}-c_{\mathbf{k}\downarrow}^{\dagger}c_{\mathbf{k}^{\prime}\downarrow}}{2}\right.\nonumber\\
& \left.+J_\perp \left(S^{+}c_{\mathbf{k}\downarrow}^{\dagger}c_{\mathbf{k}^{\prime}\uparrow}+S^{-}c_{\mathbf{k}\uparrow}^{\dagger}c_{\mathbf{k}^{\prime}\downarrow}\right)\right],\label{H_K}
\end{align}
characterized by antiferromagnetic exchange couplings $J_\parallel>0$ and $J_\perp>0$, which respectively describe classical and quantum (i.e., spin flip) processes.  The Hamiltonians $H_\text{s-d}$ and $H_\text{SC}$ could be in principle  generalized to consider the more realistic case of many superconducting bands (as in the case of Pb), but here for clarity in the presentation we only consider a single band.


Finally, the single-ion magnetic anisotropy term
\begin{equation}
H_{\text{anis}}=-D\left(S^{z}\right)^{2}, \label{Hanis}
\end{equation}
describes a spin-$S$  impurity with uniaxial anisotropy along the $\hat{z}$ direction. Physically, the case of easy-axis anisotropy ($D>0$)  favours the maximal $S^z$ projections, i.e.,  $m=\pm S$, while the hard-axis case ($D<0$)  favours an impurity ground state with projection $m=0$ for $S$ integer, or $m=\pm1/2$ for $S$ half-integer.

In a series of seminal papers, Yu, Shiba and Rusinov independently studied the above Hamiltonian in the classical limit $S\rightarrow \infty$,  $J_\parallel \rightarrow 0$, such that the dimensionless coupling parameter 
\begin{align}
\alpha_\parallel&=\frac{1}{2}J_\parallel S\rho_{0}\pi, 
\end{align}
(where $\rho_0$ the density of states at the Fermi energy in the normal state) is finite\cite{Yu65_YSR_states, Shiba_1968, Rusinov69_YSR_states}. The spin-flip term proportional to $J_\perp$ can be neglected in this limit, and an effectively single-particle Hamiltonian is obtained, describing a superconductor with a point-like Zeeman term which can always be assumed to point along the $z$ axis. In the limit of an infinitely wide conduction band, the Shiba states  are located at energies\cite{Balatsky_2006,Heinrich18_Review_single_adsorbates,Yu65_YSR_states, Shiba_1968, Rusinov69_YSR_states}
\begin{equation}
\frac{E_\text{cl}}{\Delta}=\pm\frac{1-\alpha_\parallel^{2}}{1+\alpha_\parallel^{2}},
\label{ShibaLimit}
\end{equation}
where the $\pm$ sign means that they are symmetrically located around the Fermi energy due to the electron-hole symmetry of the BCS Hamiltonian (\ref{HSC}) (here the subscript ``cl''  stands for classical approximation). 

A Shiba state can be interpreted as a discrete fermionic transition between the  many-body ground state and the first excited many-body  state, which necesarily must belong to different fermion-parity subspaces\cite{Sakurai70}. As the limit  $\alpha_\parallel \rightarrow 1$ is approached, these states become closer in energy and approach the Fermi level. Eventually, when $\alpha_\parallel =1$  they become exactly degenerate, and the system experiences a parity- and spin-changing QPT, which is signalled by the crossing of the Shiba states. In the phase with $\alpha_\parallel >1$, the effective local Zeeman potential induced by the impurity becomes strong enough to bind an extra electron, thus changing the fermionic parity in the ground state\cite{Sakurai70}. 

In addition to the classical approximation, the attempts to introduce quantum effects initiated a long time ago with the implementation of perturbative approaches\cite{Soda67_sd_exchange_in_a_SC}. Later, Zittartz and M{\"u}ller-Hartmann\cite{Zittartz_1970_I, Zittartz_1970_III} studied a Kondo impurity in a superconducting host, adapting Nagaoka's decoupling scheme of equations of motion for the Green's function\cite{Nagaoka_1965} to the superconducting case, and recovered the Shiba states in the presence of quantum fluctuations. They realized that the relevant condition leading to the QPT is actually $\Delta \approx T_K$\cite{Zittartz_1970_I, Zittartz_1970_III}. The development of the Wilson's NRG technique allowed to obtain a detailed description of the full many-body problem, and to precisely obtain the critical ratio $T^c_K/\Delta$ (e.g., $T^c_K/\Delta\simeq 0.27$ for a $S=1/2$ impurity). When $\Delta>T^c_K$, the ground state is an unscreened many-body state with total spin $S_\text{T}=S$ and even fermion parity. For $\Delta<T^c_K$, the ground state is a Kondo screened state with $S_\text{T}=S-1/2$ and odd fermionic parity \cite{Satori92_Magnetic_impurities_in_SC_with_NRG, Sakai93_Magnetic_impurities_in_SC_with_NRG,Yoshioka98_Kondo_impurity_in_SC_with_NRG,Yoshioka00_NRG_Anderson_impurity_on_SC}.

\begin{figure}[t]
\includegraphics[width=1.0\columnwidth]{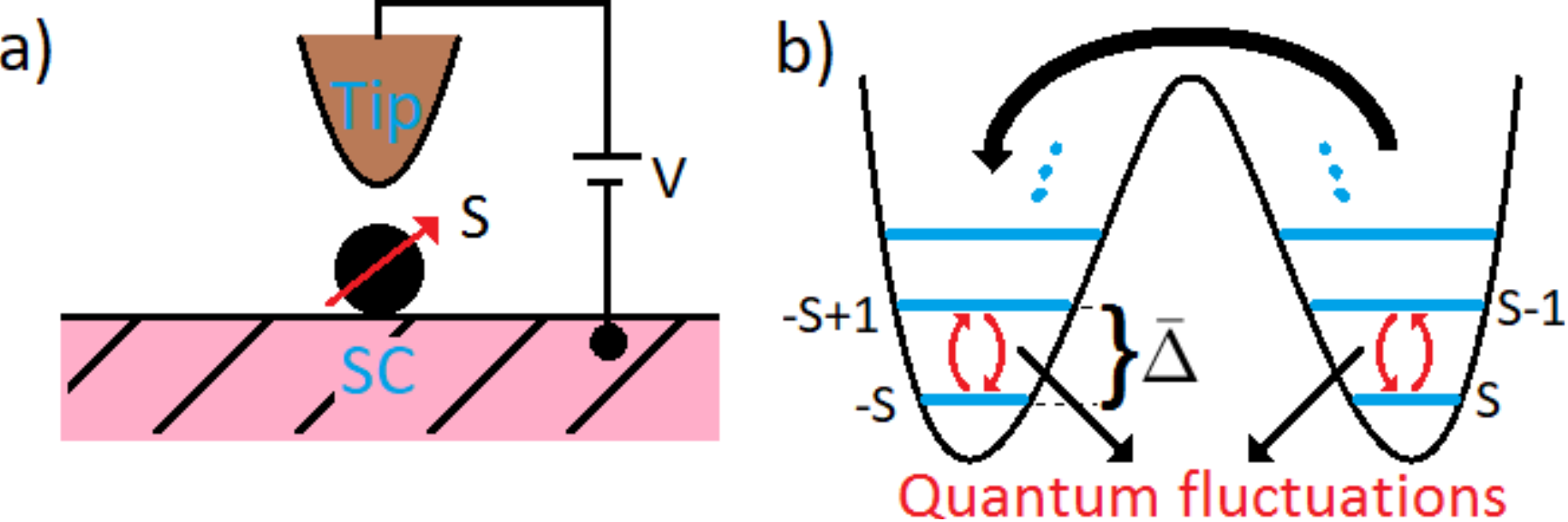}
\caption{(Color online) a) Diagram of the experimental setup. b) Level spectrum for easy-axis anisotropy ($D>0$). This figure shows the spin fluctuations from the ground state of the impurity (states $m=\pm S$) to the first excited state ($m=\pm(S-1)$) at zero temperature. Here we defined $\bar{\Delta}\equiv D(2S-1)$, see text.}
\label{Fig:Diagram}
\end{figure}

\section{Equations of motion and decoupling scheme}\label{sec:decoupling_eom}

We now outline our theoretical framework. We define the fermionic propagators in imaginary time $0\leq \tau <\beta$ (with $\beta=1/T$ since we are taking units where $k_B=1$)\cite{fetter}
\begin{eqnarray}
g_{\mathbf{k},\mathbf{k}^\prime}\left(\tau-\tau^{\prime}\right) & = & -\langle T_{\tau}c_{\mathbf{k}\uparrow}\left(\tau\right)c_{\mathbf{k}^\prime\uparrow}^{\dagger}\left(\tau^{\prime}\right)\rangle \label{g}\\
f_{\mathbf{k},\mathbf{k}^\prime}\left(\tau-\tau^{\prime}\right) & =&-\langle T_{\tau}c_{\mathbf{k}\uparrow}\left(\tau\right)c_{-\mathbf{k}^\prime\downarrow}\left(\tau^{\prime}\right)\rangle \label{f}\\
\bar{f}_{\mathbf{k},\mathbf{k}^\prime}\left(\tau-\tau^{\prime}\right) & =&-\langle T_{\tau}c_{-\mathbf{k}\downarrow}^{\dagger}\left(\tau\right)c_{\mathbf{k}^\prime\uparrow}^{\dagger}\left(\tau^{\prime}\right)\rangle \label{fbar}\\
\bar{g}_{\mathbf{k},\mathbf{k}^\prime}\left(\tau-\tau^{\prime}\right) & =&-\langle T_{\tau}c_{-\mathbf{k}\downarrow}^{\dagger}\left(\tau\right)c_{-\mathbf{k}^\prime\downarrow}\left(\tau^{\prime}\right)\rangle \label{gbar}.
\end{eqnarray}
where  $T_\tau$ is the imaginary-time ordering operator. Here, $g_{\mathbf{k},\mathbf{k}^\prime}\left(\tau-\tau^\prime \right)$ and $f_{\mathbf{k},\mathbf{k}^\prime}\left(\tau-\tau^\prime \right)$ are, respectively, the normal and anomalous fermionic correlators. Using the SU(2) symmetry of our model (\ref{eq:H_total-1}), we have dropped the spin indices in Eqs. (\ref{g})-(\ref{fbar}), as the correlators satisfy the relations
\begin{align}
-\langle T_{\tau}c_{\mathbf{k}\uparrow}\left(\tau\right)c_{\mathbf{k}^\prime\uparrow}^{\dagger}\left(\tau^{\prime}\right)\rangle &=-\langle T_{\tau}c_{\mathbf{k}\downarrow}\left(\tau\right)c_{\mathbf{k}^\prime\downarrow}^{\dagger}\left(\tau^{\prime}\right)\rangle\\
-\langle T_{\tau}c_{\mathbf{k}\uparrow}\left(\tau\right)c_{-\mathbf{k}^\prime\downarrow}\left(\tau^{\prime}\right)\rangle &= -\langle T_{\tau}c_{\mathbf{k}\downarrow}\left(\tau\right)c_{-\mathbf{k}^\prime\uparrow}\left(\tau^{\prime}\right)\rangle,
\end{align}
Introducing the Fourier representation 
\begin{align}
g_{\mathbf{k},\mathbf{k}^\prime}\left(i\nu_n\right)&=\llangle c_{\mathbf{k}\uparrow};c^\dagger_{\mathbf{k}^\prime\uparrow} \rrangle\left(i\nu_n\right),\nonumber\\
&=\int_0^\beta d\tau\ e^{-i\nu_n\left(\tau-\tau^{\prime}\right)} g_{\mathbf{k},\mathbf{k}^\prime}\left(\tau-\tau^{\prime}\right),
\end{align} 
[and similarly for the other correlators (\ref{f})-(\ref{gbar})] where $\nu_{n}=\pi(2n+1)/\beta$ are the fermionic Matsubara frequencies, we can compactly express the fermionic Green's function using the Nambu-matrix notation: 
\begin{eqnarray}
\mathbf{G}\left(i\nu_n\right)&=&\left(\begin{array}{cc}
g\left(i\nu_{n}\right) & f\left(i\nu_{n}\right)\\
\bar{f}\left(i\nu_{n}\right) & \bar{g}\left(i\nu_{n}\right)
\end{array}\right),\nonumber\\
&=&\frac{1}{V}\sum_{\mathbf{k},\mathbf{k}^\prime}\left(\begin{array}{cc}
g_{\mathbf{k},\mathbf{k}^\prime}\left(i\nu_n\right) & f_{\mathbf{k},\mathbf{k}^\prime}\left(i\nu_n\right)\\
\bar{f}_{\mathbf{k},\mathbf{k}^\prime}\left(i\nu_n\right) & \bar{g}_{\mathbf{k},\mathbf{k}^\prime}\left(i\nu_n\right)
\end{array}\right), 
\label{G}
\end{eqnarray}
In particular, the knowledge of the local correlator $g\left(i\nu_{n}\right)$ allows to obtain, upon analytical continuation to real frequencies $i\nu_{n}\rightarrow \omega +i\epsilon$, the local density of states 
\begin{align}
\rho\left(\omega\right)&\xrightarrow[\epsilon\rightarrow 0]{}-\frac{1}{\pi}\text{Im}\left[g\left(\omega + i\epsilon \right)\right],
\end{align}
which is related to the STM differential conductace $dI/dV\left(\omega\right)$ at the position of the impurity (see e.g., Ref. \onlinecite{Fisher07_RMP_STM_HTSC}).
  
The individual components  of the Nambu matrix Eq. (\ref{G}) satisfy the relation 
\begin{align}
\left[\mathbf{G}\right]_{AB}\left(z\right)&=-\left[\mathbf{G}\right]_{BA}\left(-z\right),\label{symmetries}
\end{align}
where $A,B$ are the fermionic operators entering the definitions (\ref{g})-(\ref{gbar})\cite{Zittartz_1970_I, Zittartz_1970_III}.

In the absence of the $H_\text{s-d}$ term, the unperturbed propagators have a closed analytical form, which can be expressed as \cite{Balatsky_2006,fetter,Tinkham_Introduction_to_superconductivity}
\begin{eqnarray}
g^{\left(0\right)}_{\mathbf{k},\mathbf{k}^\prime}\left(i\nu_{n}\right)&=&\delta_{\mathbf{k},\mathbf{k}^\prime}\left[\frac{u_{\mathbf{k}}^{2}}{i\nu_{n}-E_{\mathbf{k}}}+\frac{v_{\mathbf{k}}^{2}}{i\nu_{n}+E_{\mathbf{k}}}\right] \label{g0}\\
f^{\left(0\right)}_{\mathbf{k},\mathbf{k}^\prime}\left(i\nu_{n}\right)&=&\delta_{\mathbf{k},\mathbf{k}^\prime}\left[\frac{2u_{\mathbf{k}}v_{\mathbf{k}}}{i\nu_{n}-E_{\mathbf{k}}}-\frac{2u_{\mathbf{k}}v_{\mathbf{k}}}{i\nu_{n}+E_{\mathbf{k}}}\right], \label{f0}\end{eqnarray}
where $u_{\mathbf{k}}^{2}=\frac{1}{2}\left(1+\frac{\varepsilon_{\mathbf{k}}}{E_{\mathbf{k}}}\right)$ and $v_{\mathbf{k}}^{2}=\frac{1}{2}\left(1-\frac{\varepsilon_{\mathbf{k}}}{E_{\mathbf{k}}}\right)$are, respectively, the electron and hole weights of the  Bogoliubov quasiparticle $\gamma_{\mathbf{k},\uparrow}=u_\mathbf{k} c_{\mathbf{k},\uparrow} + v_\mathbf{k} c^\dagger_{-\mathbf{k},\downarrow}$, and 
$E_{\mathbf{k}}=\sqrt{\varepsilon_{\mathbf{k}}^{2}+\Delta^{2}}$ its eigenenergy \cite{fetter,Tinkham_Introduction_to_superconductivity}. The delta function $\delta_{\mathbf{k},\mathbf{k}^\prime}$ appears since in the absence of the $H_\text{s-d}$ term, the translational symmetry in the 2D plane is preserved and therefore the 2D quasi-momentum $\mathbf{k}$ is a conserved quantity.

Replacing Eqs.  (\ref{g0}) and (\ref{f0}) into Eq. (\ref{G}), and performing the sum over momenta as an integral over a flat conduction-band, i.e $\frac{1}{V}\sum_\mathbf{k} \rightarrow \rho_0 \int_{-W}^{W}d\epsilon$,  the analytical expression of the unperturbed local propagator is
\begin{align}
\mathbf{G}^{\left(0\right)}\left(i\nu_n\right)	&=	\left(\begin{array}{cc}
g^{\left(0\right)}\left(i\nu_n\right) & f^{\left(0\right)}\left(i\nu_n\right)\\
\bar{f}^{\left(0\right)}\left(i\nu_n\right) & \bar{g}^{\left(0\right)}\left(i\nu_n\right)
\end{array}\right),\nonumber\\
	&=	\frac{-2\rho_{0}\tan^{-1}\left(\frac{W}{\sqrt{\Delta^{2}-\left(i\nu_n\right)^{2}}}\right)}{\sqrt{\Delta^{2}-\left(i\nu_n\right)^{2}}}\left(\begin{array}{cc}
i\nu_n & \Delta\\
\Delta & i\nu_n
\end{array}\right),\label{G0}
\end{align}
with $W$ half the bandwidth related to $\rho_0$ through the normalization condition $\rho_0=1/2W$.

The first step to obtain the Nambu Green's function $\mathbf{G}\left(i\nu_n\right)$ in the presence of the term $H_\text{s-d}$ is to obtain the equation of motion for the fermionic operator $\partial_\tau c_{\mathbf{k},\sigma}\left(\tau\right)=\left[H,c_{\mathbf{k},\sigma}\left(\tau\right)\right]$. Replacing this result in the definitions (\ref{g})-(\ref{gbar}) and passing to Matsubara-frequency representation, we obtain the expression\cite{Zubarev_1960, Costi00,Zittartz_1970_I, Nagaoka_1965}: 
\begin{align}
\mathbf{G}\left(i\nu_{n}\right) & =\mathbf{G}^{\left(0\right)}\left(i\nu_{n}\right)+\mathbf{G}^{\left(0\right)}\left(i\nu_{n}\right)\mathbf{T}\left(i\nu_{n}\right)
\mathbf{G}^{\left(0\right)}\left(i\nu_{n}\right),\label{G_tmatrix}
\end{align}
where
\begin{widetext}
\begin{align}
\mathbf{T}\left(i\nu_{n}\right)\equiv\frac{J_\parallel^2}{4V}\sum_{\mathbf{q},\mathbf{q}^\prime}
\left(\begin{array}{cc}
\llangle  S^zc_{\mathbf{q}\downarrow};S^zc_{\mathbf{q}^\prime\downarrow}^{\dagger}\rrangle(i\nu_n) & \llangle S^zc_{\mathbf{q}\downarrow};S^zc_{\mathbf{q}^\prime\uparrow}\rrangle(i\nu_n)\\
\llangle  S^zc_{\mathbf{q}\uparrow}^{\dagger};S^zc_{\mathbf{q}^\prime\downarrow}^{\dagger}\rrangle(i\nu_n) & \llangle  S^zc_{\mathbf{q}\uparrow}^{\dagger};S^zc_{\mathbf{q}^\prime\uparrow}\rrangle(i\nu_n) \end{array}\right)
+\frac{J_\perp^2}{4V}\sum_{\mathbf{q},\mathbf{q}^\prime}\left(\begin{array}{cc}
\llangle  S^{-}c_{\mathbf{q}\downarrow};S^{+}c_{\mathbf{q}^\prime\downarrow}^{\dagger}\rrangle(i\nu_n) & \llangle S^{-}c_{\mathbf{q}\downarrow};S^{+}c_{\mathbf{q}^\prime\uparrow}\rrangle(i\nu_n)\\
\llangle  S^{-}c_{\mathbf{q}\uparrow}^{\dagger};S^{+}c_{\mathbf{q}^\prime\downarrow}^{\dagger}\rrangle(i\nu_n) & \llangle  S^{-}c_{\mathbf{q}\uparrow}^{\dagger};S^{+}c_{\mathbf{q}^\prime\uparrow}\rrangle(i\nu_n) \end{array}\right).\label{t_matrix}
\end{align}
is the $t-$matrix of the problem, which contains all the effects of the magnetic impurity. This expression is formally exact, provided we know the exact form of the correlators 
\begin{align}
\llangle  S^{a}\eta_{\mathbf{q}\sigma};S^{b}\eta^\prime_{\mathbf{q}^\prime\sigma^\prime}\rrangle(i\nu_n)&=
\int_0^\beta d\tau\ e^{-i\nu_n\left(\tau-\tau^\prime\right)} \langle T_\tau S^{a}\left(\tau\right)\eta_{\mathbf{q}\sigma}\left(\tau\right) S^{b}\left(\tau^{\prime}\right)\eta^\prime_{\mathbf{q}^\prime\sigma^\prime}\left(\tau^{\prime}\right)\rangle,\label{correlator_t_matrix}
\end{align} 
\end{widetext}
with $S^{a (b)}=\left\{S^+, S^-\ \text{or}\ S^z\right\}$, and $\eta_{\mathbf{q}\sigma}=\left\{c_{\mathbf{q}\sigma}\ \text{or}\ c^\dagger_{\mathbf{q}\sigma}\right\}$. However, due to the many-body nature of the problem, an infinite hierarchy of higher-order correlators must be known to have a closed expression for Eq. (\ref{correlator_t_matrix})\cite{Nagaoka_1965}. Therefore, in order to make progress, a truncation of this hierarchy of correlators must be introduced. Here we propose the following approximate decoupling:
\begin{align}
&\langle T_\tau S^{a}\left(\tau\right)\eta_{\mathbf{q}\sigma}\left(\tau\right) S^{b}\left(\tau^{\prime}\right)\eta^\prime_{\mathbf{q}^\prime\sigma^\prime}\left(\tau^{\prime}\right)\rangle \nonumber \\& \approx \langle T_\tau S^{a}\left(\tau\right)S^{b}\left(\tau^{\prime}\right)\rangle \times \langle T_\tau \eta_{\mathbf{q}\sigma}\left(\tau\right) \eta^\prime_{\mathbf{q}^\prime\sigma^\prime}\left(\tau^{\prime}\right)\rangle,\label{decoupling}
\end{align}
which is valid in the weak-coupling limit $\{\rho_0 J_\parallel,\rho_0 J_\perp \} \rightarrow 0$, i.e., when both spin and electron subsystems evolve more or less independently. Consequently, in the rest of this work we will focus on this regime of parameters, which corresponds to the ``unscreened'' region of the quantum phase diagram $T_K\ll \Delta$. Due to this limitation,  Kondo correlations cannot be recovered within our approach. Nevertheless, the decoupling  (\ref{decoupling}) is still very useful, as it allows to close the set of equations of motion and to obtain an integral equation for the Nambu Green's function $\mathbf{G}\left(i\nu_n\right)$ with meaningful information about spin fluctuations in the presence of anisotropy. 

Although our approach shares some similarities with Nagaoka's decoupling method\cite{Nagaoka_1965,Zittartz_1970_I, Zittartz_1970_III}, there is a crucial difference regarding the spin degrees of freedom: in Nagaoka's method, the product of spin operators are considered only at the level of static averages with no intrinsic dynamics, whereas in our case the spin correlators retain their dynamics, including the time evolution dictated by the anisotropy term $H_\text{anis}$. 

We note that the full (i.e., ``dressed'') spin correlators are in principle needed in Eq. (\ref{decoupling}). This introduces an additional set of equations of motion for the spin correlators, which must be obtained in order to obtain the electronic propagator, complicating the application of the method. However, within the weak-coupling regime, it is rather natural to replace the full spin correlators by the unperturbed ones, i.e.: 
\begin{align}
\left\langle T_\tau S^{-}(\tau)S^{+}(\tau^{\prime})\right\rangle&\approx \left\langle T_\tau S^{-}(\tau)S^{+}(\tau^{\prime})\right\rangle_0,\nonumber\\
\left\langle T_\tau S^{z}(\tau)S^{z}(\tau^{\prime})\right\rangle&\approx \left\langle T_\tau S^{z}(\tau)S^{z}(\tau^{\prime})\right\rangle_0, \label{approx_spin_correlators}
\end{align}
which can be computed analytically and allows to simplify the problem. Using the Matsubara-frequency representation of the spin correlators
\begin{align}
\llangle S^{a};S^{b}\rrangle_0 \left(i\omega_{l}\right)&=\int_0^\beta d\tau \ e^{i\omega_l \left(\tau-\tau^\prime\right)}\ \left\langle T_\tau S^{a}(\tau)S^{b}(\tau^\prime)\right\rangle_0,\label{Fourier_SaSb}
\end{align}
where $\omega_l=2\pi l/\beta$ are the bosonic Matsubara frequencies, the unperturbed $\llangle S^{z};S^{z}\rrangle_0 \left(i\omega_{l}\right)$ correlator is
\begin{align}
\llangle S^{z};S^{z}\rrangle_0 \left(i\omega_{l}\right) & =\beta\delta_{\omega_l,0}\langle \left( S^{z} \right)^2\rangle_0.\label{eq:szsz}
\end{align}
This is a static quantity since the operator $S^z$ commutes with $H_\text{anis}$, and therefore is conserved in the absence of $H_\text{s-d}$. The thermodynamical average is easily computed as
\begin{align}
\langle \left( S^{z} \right)^2\rangle_0=\sum_{m=-S}^{S}\frac{m^{2}e^{-\beta \Delta^0_{m}}}{Z_0}, \label{sz2}
\end{align}
where we have defined $\Delta^0_{m} \equiv E^0_{m}-E^0_\text{min}$, with $E^0_m\equiv -D m^2$ the eigenvalues of $H_\text{anis}$, and $E^0_\text{min}\equiv \text{min} \{E^0_m\}$. The quantity $Z_0=\sum_{m=-S}^{S}e^{-\beta \Delta^0_{m}}$ is the spin partition function, computed up to an irrelevant prefactor.  

The (unperturbed) dynamical correlator
\begin{align}
\llangle S^{-};S^{+}\rrangle_0 \left(i\omega_{l}\right) & =\sum_{m=-S}^{S}\frac{A_{m}}{i\omega_{l}-\left(\Delta^0_{m+1}-\Delta^0_{m}\right)},\label{eq:smsp}
\end{align}
is obtained introducing the Heisenberg representation $S^{\pm}\left(\tau \right)=e^{H \tau} S^{\pm} e^{-H \tau}$ and the identity operator $\mathbf{1}=\sum_{m^\prime=-S}^S|m^\prime\rangle \langle m^\prime |$ in Eq. (\ref{Fourier_SaSb}), and is a crucial quantity in this work since it encodes the information about the quantum fluctuations in the system. In Eq. (\ref{eq:smsp}) we have defined the matrix element
\begin{align*}
A_{m} & \equiv\frac{S\left(S+1\right)-m\left(m+1\right)}{Z_0}\left(e^{-\beta\Delta^0_{m+1}}-e^{-\beta\Delta^0_{m}}\right).
\end{align*}

Returning to Eq. (\ref{t_matrix}), and implementing the decoupling (\ref{decoupling}) along with the approximation (\ref{approx_spin_correlators}), the $t-$matrix compactly writes as
\begin{align}
\mathbf{T}\left(i\nu_{n}\right)&=\frac{1}{4{\beta}}\sum_{\omega_l=-\infty}^\infty \mathbf{G}\left(i\nu_n-i\omega_l\right)  \left[J_\parallel^2 \beta \delta_{\omega_l,0} \langle \left(S^{z}\right)^{2}\rangle_0 \right. \nonumber\\ &  \left. +J_\perp^2 \llangle S^{-};S^{+}\rrangle_0 \left(i\omega_{l}\right)\right],\label{t_matrix_approx}
\end{align}
i.e., the $t-$matrix is obtained as a convolution of the electronic Green's function and the dynamical spin correlators. Therefore, Eq. (\ref{t_matrix_approx}) toghether with (\ref{G_tmatrix}) describe an integral equation for the Nambu Green's function $\mathbf{G}\left(i\nu_n \right)$

\begin{align}
\mathbf{G}\left(z\right) & =\mathbf{G}^{\left(0\right)}\left(z\right)+\mathbf{G}^{\left(0\right)}\left(z\right)
\frac{1}{4{\beta}}\sum_{\omega_l=-\infty}^\infty \mathbf{G}\left(z-i\omega_l\right) \nonumber\\
&\times \left[J_\parallel^2 \beta \delta_{\omega_l,0} \langle \left(S^{z}\right)^{2}\rangle_0 +J_\perp^2 \llangle S^{-};S^{+}\rrangle_0 \left(i\omega_{l}\right)\right]
\mathbf{G}^{\left(0\right)}\left(z\right),\label{G_main_equation}
\end{align}
where we have performed the analytic continuation to complex frequencies $i\nu_{n}\rightarrow z$. 
In principle, the $\left(2\times 2\right)$-matrix structure of Eq. (\ref{G_main_equation}) implies  solving a set of 4 coupled integral equations. However, using the Nambu symmetries of the problem, detailed in Eq. (\ref{symmetries}), along with the SU(2) symmetry and the intrinsic particle-hole symmetry of the band $\varepsilon_\mathbf{k}$, the number of unknown functions can be reduced from 4 to 2 
[i.e, particle-hole symmetry implies that $g\left(z\right)= \bar{g}\left(z\right)$ and $f\left(z\right)= \bar{f}\left(z\right)$]. The remaining equations for the Green's functions $g\left(z\right)$ and $f\left(z\right)$ can be further decoupled using the change of variables $g_{\pm}\left(z\right) =\frac{1}{2}\left(g\left(z\right)\pm f\left(z\right)\right)$\cite{Zittartz_1970_I}. In terms of the dimensionless functions $\tilde{g}_\pm\left(z\right)=g_\pm\left(z\right)/\rho_0$, we obtain two decoupled scalar equations
\begin{widetext}
\begin{align}
\tilde{g}_{\pm}\left(z\right) & =\tilde{g}_{\pm}^{\left(0\right)}\left(z\right)+\frac{4\left(\tilde{g}_{\pm}^{\left(0\right)}\left(z\right)\right)^{2}}{\pi^2}\left[ \frac{\alpha_{\parallel}^{2}\langle \left(S^{z}\right)^{2}\rangle_0}{S^2} \tilde{g}_{\pm}\left(z\right) +\frac{\alpha_{\perp}^2}{ S^2 \beta}\sum_{\omega_l=-\infty}^\infty 
\sum_{m=-S}^{S}\frac{A_{m}}{i\omega_{l}-\left(\Delta^0_{m+1}-\Delta^0_{m}\right)}\tilde{g}_{\pm}\left(z-i\omega_{l}\right)\right].\label{eq:eq_integral}
\end{align}
\end{widetext}
where we have introduced  the dimensionless couplings
\begin{align}
\alpha_{\parallel} & \equiv\frac{J_{\parallel}\rho_{0}S\pi}{2}, &\alpha_{\perp} & \equiv\frac{J_{\perp}\rho_{0}S\pi}{2}.\label{dimensionless_alpha}
\end{align}
Eq. (\ref{eq:eq_integral}) is one of the most important results in this work. It can be interpreted as a  generalization of the classical-spin approximation \cite{Balatsky_2006, Yu65_YSR_states, Shiba_1968, Rusinov69_YSR_states} that incorporates  the effects of quantum fluctuations and anisotropy in the weak-coupling limit. One can easily check that in absence of the spin-flip term (i.e., $\alpha_\perp=0$), Eq. (\ref{eq:eq_integral}) indeed reduces to the classical-spin limit which can be solved analytically\cite{Balatsky_2006}. In fact, if we assume  easy-axis anisotropy $D>0$, and $T\rightarrow 0$, the spin operator $S^z$ acquires the classical value $\langle \left(S^{z}\right)^{2}\rangle_0\rightarrow S^2$, and we obtain
\begin{align}
\tilde{g}_{\pm}\left(z\right) & =\frac{\tilde{g}_{\pm}^{\left(0\right)}\left(z\right)}{1-\frac{4\alpha_{\parallel}^{2}}{\pi^2}\left(\tilde{g}_{\pm}^{\left(0\right)}\left(z\right)\right)^{2}}.
\end{align}
The poles of this equation allows to recover the position of the classical Shiba states Eq. (\ref{ShibaLimit}).

We now solve Eq. (\ref{eq:eq_integral}) in the presence of quantum fluctuations, which is the most interesting case for our purposes. The solution in this case is complicated by the fact that $\tilde{g}_\pm\left(z\right)$ is non-diagonal in the frequency domain, a consequence of the dynamical nature of the spin correlator (\ref{eq:smsp}) and, ultimately, a consequence of the many-body nature of the problem. We first note that for a fixed value  of $z=\omega + i \epsilon$, where $\epsilon >0 $ is infinitesimally small, the imaginary axis can be discretized according to the Matsubara frequencies as $z_k=z-i\omega_k$. Then, defining  a vector  $\mathbf{g}$, whose $k$-th element is $\mathbf{g}_k\equiv \tilde{g}_{\pm}\left(z_k\right)$, Eq. (\ref{eq:eq_integral}) can be compactly expressed in matrix form as 
\begin{align}
\mathbf{g}_k&=\mathbf{g}^{\left(0\right)}_k + \sum_{l=-\infty}^\infty \mathbf{K}_{kl}. \mathbf{g}_l,\label{matrix_equation}
\end{align} 
where the elements of the infinite matrix $\mathbf{K}$ are defined as 
\begin{align}
\mathbf{K}_{kl}&=\frac{4\left(\mathbf{g}_{k}^{\left(0\right)} \right)^{2}}{\pi^2}\left[\frac{\alpha_{\parallel}^{2}\langle \left(S^{z}\right)^{2}\rangle_0}{S^2}\delta_{kl} \right.\nonumber \\
&\left. +\frac{\alpha_{\perp}^{2}}{S^2 \beta} \sum_{m=-S}^{S}\frac{A_{m}}{i\omega_{l}-i\omega_{k}-\left(\Delta^0_{m+1}-\Delta^0_{m}\right)} \right].
\end{align}
Although Eq. (\ref{matrix_equation}) seems formally simple, the infinite rank of the matrix $\mathbf{K}$ represents a technical complication that must be addressed previously to attempt a numerical solution of this problem. To this end, we make use of the asymptotic property of physical Green's functions \cite{fetter}
\begin{align}
\tilde{g}_{\pm}\left(z\right) &\xrightarrow[|z|\rightarrow \infty]{} \frac{W}{z}\label{asymptotic_behavior},
\end{align}
to introduce a truncation in (\ref{matrix_equation}). We therefore split the sum into $\sum_{l=-\infty}^\infty =\sum_{l=-l_\text{max}}^{l_\text{max}} +\sum_{l_\text{max}<|l|}$, where $l_\text{max}$ is an integer chosen such that for $|\omega_l|>\omega_{l_\text{max}}$ we can safely use Eq.  (\ref{asymptotic_behavior}). Then, we can  write
\begin{align}
\mathbf{g}_k&\approx \mathbf{g}^{\left(0\right)}_k + \mathbf{S}_k+ \sum_{l=-l_\text{max}}^{l_\text{max}} \mathbf{K}_{kl} \mathbf{g}_l, \label{matrix_equation_2}
\end{align}
where we constrain the values of  $i\omega_k$ on the left hand-side to be in the range $-i\omega_{l_\text{max}}<i\omega_k<i\omega_{l_\text{max}}$.  In this form, the original Eq. (\ref{eq:eq_integral}) can be expressed in terms of a {\it{finite}} matrix $\mathbf{K}$ of size $l_\text{max} \times l_\text{max}$, and a additional vector 
\begin{align}
\mathbf{S}_k&= \frac{4\alpha_{\perp}^{2}\left(\mathbf{g}_{k}^{\left(0\right)} \right)^{2}}{\pi^2S^2} \sum_{m=-S}^{S} \frac{1}{\beta}\sum_{l_\text{max}<|l|} \frac{W}{\left(z-i\omega_l\right)}\nonumber\\
&\times\frac{A_{m}}{i\omega_{l}-i\omega_{k}-\left(\Delta^0_{m+1}-\Delta^0_{m}\right)}, \label{matrix_equation_3}
\end{align}
that contains the contribution of all the higher Matsubara frequencies, and which can be evaluated analytically in terms of the digamma functions $\Psi\left(z\right)$ [see Eq. (\ref{final_Sk}) in the Appendix].

We solve Eq. (\ref{matrix_equation_2}) numerically implementing the $LU$-decomposition, and once we obtain the vector $\mathbf{g}$ (computed for each value of the real frequency $\omega$ in $z=\omega + i\epsilon$), we extract the element $\mathbf{g}_0=\tilde{g}_\pm\left(\omega + i\epsilon \right)$ to obtain the local density of states  at the position of the impurity:
\begin{align}
\tilde{\rho}\left(\omega\right)&=-\frac{1}{\pi}\text{Im}\left[ \tilde{g}_+\left(\omega + i\epsilon \right)+\tilde{g}_-\left(\omega + i\epsilon \right)\right].\label{eq:rho_tilde}
\end{align}

In practice, the asymptotic behavior Eq. (\ref{asymptotic_behavior})  is reached within acceptable error levels in (\ref{matrix_equation_2}) choosing a Matsubara-frequency cutoff $\omega_{l_\text{max}}\approx 100 \Delta$. We have checked that increasing this cutoff does not significatively modify our results. From here, it follows that the value $l_\text{max}$ is determined by the relation 
\begin{align}
l_\text{max}=\frac{100 \Delta}{2 \pi T},
\end{align}
which means that the size of the matrix $\mathbf{K}$ explicitly depends on the ratio $\Delta/T$. This feature imposes a minimal temperature below which the numerical solution of (\ref{matrix_equation_2}) is beyond our current computational power. At the same time, as mentioned in Sec. \ref{sec:model} we must restrict the temperature to the regime $T\lesssim T_c/2$, in order to avoid the extra self-consistency step implied in the BCS gap-equation at finite temperatures \cite{Tinkham_Introduction_to_superconductivity}. A ``sweet spot'' where  a good compromise between these two limits is obtained is $0.05\Delta \lesssim T\lesssim0.5\Delta$. Taking the experimental value for the superconducting gap in Pb as $\Delta\simeq 1.3$meV from Ref. \onlinecite{Hatter2017_Scaling_of_YSR_energies}, we obtain the temperature range 0.87 K $<T<$ 8.7 K, which is consistent with the experimental temperature $T_\text{exp}\simeq 1.2$ K used in that reference. This is also a realistic range of temperatures in similar experiments\cite{Menard2015,Kezilebieke2018,Ji08_YSR_states,Hatter_2015,Hatter2017_Scaling_of_YSR_energies,Heinrich_2015,Ruby_2015,Choi_2017}.

\section{Results}\label{sec:results_T0}
From the considerations in previous sections, we expect that the anisotropy $D$ profoundly modifies the behavior of the quantum impurity. In the case of vanishing Kondo temperature $T_K \ll \Delta$, our results depend strongly on the ratio $D/T$. We therefore analyze separately three different regimes: the case $D\ll T$, the case  $D\gg T$, and the case $D \simeq T$. 
\subsection{Regime $D\ll T$}\label{subsec:DllT}
\begin{figure}[t]
\includegraphics[width=1.0\columnwidth]{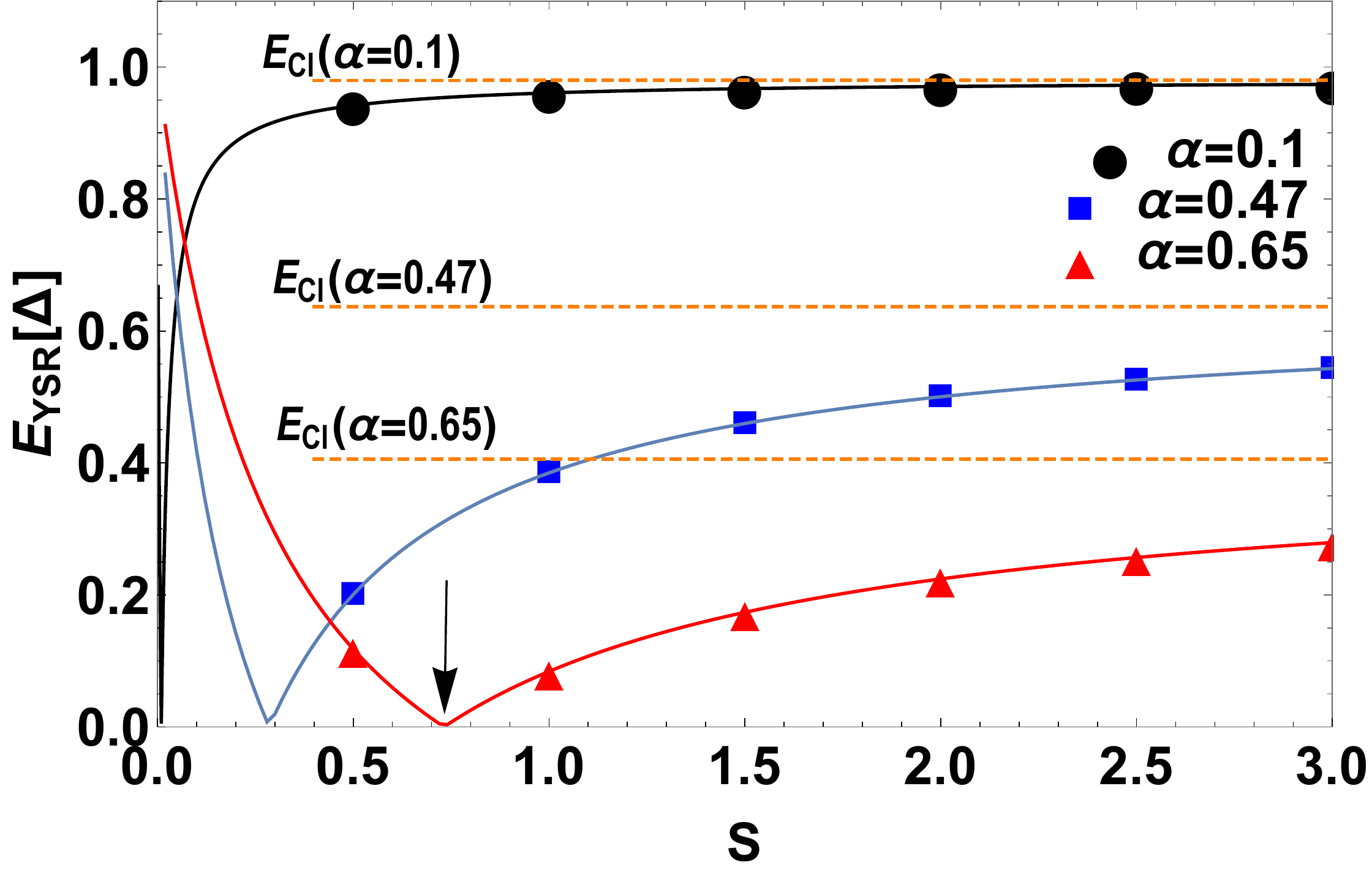}
\caption{(Color online) YSR energy $E_\text{YSR}$ as a function of the spin impurity $S$ for $\alpha=0.1$ (black points), $\alpha=0.47$ (blue squares) and $\alpha=0.65$ (red triangles). Continuous lines are a guide to the eye. The classical Shiba energy independent of $S$ are showed as orange dashed horizontal lines for the corresponding values of $\alpha$. The arrow marks the putative quantum phase transition predicted by perturbative approach.}
\label{Fig:Shiba}
\end{figure}

This regime can be either associated to a ``high temperature'' situation where all spin states are thermally occuppied (in our case, this situation should be consistent with the requirement that $T\lesssim T_c/2$), or to a vanishingly small anisotropy $D\rightarrow 0$, in which case the term $H_\text{anis}$ drops. In any of these cases, the result is the same: the SU(2) symmetry is restored, and the thermodynamical average  Eq. (\ref{sz2}) and the  dynamical correlator  Eq. (\ref{eq:smsp}) become, respectively
\begin{eqnarray}
\langle \left( S^{z} \right)^2\rangle_0&\xrightarrow[\Delta^0_m/T\rightarrow 0]{}& \frac{S(S+1)}{3},\\
\llangle S^{-};S^{+}\rrangle_0 \left(i\omega_{l}\right) & \xrightarrow[\Delta^0_m/T\rightarrow 0]{}&\frac{2S(S+1)}{3} \beta\delta_{\omega_l,0}.
\end{eqnarray}
Replacing these results in (\ref{eq:eq_integral}), the function $\tilde{g}_{\pm}\left(z\right)$ can be obtained analytically: 
\begin{align}
\tilde{g}_{\pm}\left(z\right) & =\frac{\tilde{g}_{\pm}^{0}\left(z\right)}{1-\frac{4}{\pi^{2}}\frac{S\left(S+1\right)}{S^2}\left(\frac{\alpha_{\parallel}^{2}}{3}+\frac{2\alpha_{\perp}^{2}}{3}\right)\left(\tilde{g}_{\pm}^{0}\left(\tilde{z}\right)\right)^{2}},\label{eq:Shiba_c}
\end{align}
where the poles determine the position of the Shiba states [here we have used the limit of infinite bandwidth $W\rightarrow \infty$, for consistency with Eq. (\ref{ShibaLimit})]
\begin{equation}
\frac{E_\text{YSR}}{\Delta}=\pm\frac{1-\frac{S\left(S+1\right)}{S^2}\left(\frac{\alpha_{\parallel}^{2}}{3}+\frac{2\alpha_{\perp}^{2}}{3}\right)}{1+\frac{S\left(S+1\right)}{S^2}\left(\frac{\alpha_{\parallel}^{2}}{3}+\frac{2\alpha_{\perp}^{2}}{3}\right)}.
\label{EQ_DeqZero}
\end{equation}
for $\tilde{g}_+\left(z\right)$ and $\tilde{g}_-\left(z\right)$ respectively. In the fully isotropic case $\alpha_\parallel=\alpha_\perp=\alpha$, Eq. (\ref{EQ_DeqZero}) is qualitatively similar to the classical YSR result Eq. (\ref{ShibaLimit}), with the difference that the coupling must be renormalized as $\alpha\rightarrow \tilde{\alpha}=\alpha \sqrt{\frac{S\left(S+1\right)}{S^2}}$ (see also Ref. \onlinecite{Kirsanskas15_YSR_states_in_phase_biased_SC_QD_SC_junctions}). This result is physically appealing since the factor $S\left(S+1\right)$ is the expectation value of  the operator  $\mathbf{S}^2$ in the SU(2) symmetric case. Therefore, the factor $\sqrt{\frac{S\left(S+1\right)}{S^2}}$ is a \emph{quantitative measure} of the amount of quantum fluctuations in the system: The value 1 corresponds to the classical limit $S\rightarrow \infty$, where Eq. (\ref{EQ_DeqZero}) converges to (\ref{ShibaLimit}), and any value larger than 1 can be attributed to the effect of quantum fluctuations. In Fig. \ref{Fig:Shiba} we show the position of the YSR states in the isotropic limit   $\alpha_\parallel=\alpha_\perp=\alpha$ as a function of spin $S$ for $\alpha=0.1$, $0.47$ and $0.65$, and we compare each case with the corresponding classical limit. As expected, the classical limit is only recovered when $S\rightarrow\infty$, and the lowest-order correction in the small parameter $S^{-1}$ is
\begin{equation}
\frac{E_\text{YSR}}{\Delta}\xrightarrow[S\rightarrow \infty]{}\frac{E_\text{cl}}{\Delta}-\frac{\alpha^{2}}{\left(1+\alpha^{2}\right)^{2}}\frac{1}{S}+\mathcal{O}\left(S^{-2}\right).\label{isotropic_Shiba}
\end{equation}
Note that due to the slow dependence $S^{-1}$, the classical result is recovered only for unphysically large values of $S$ in the fully isotropic case. This is consistent with recent results obtained by {\v Z}itko using NRG, who studied the quantum-to-classical crossover in impurities in superconductors (see Ref. \onlinecite{Zitko_2017}). Through a numerical fit of the NRG results, he obtained a correction term proportional to $S^{-\nu}$, with exponent $\nu=1.1$, very close to our analytical result. 

Taking $S$ as a continuous variable, Eq. (\ref{isotropic_Shiba}) predicts a QPT as a function of $S$, which occurs at the critical value $S_c=\alpha^{2}/(1-\alpha^{2})$ (see Fig. \ref{Fig:Shiba}). This is similar to Ref. \onlinecite{Zitko_2017}, where a critical value $S_c$ is found with NRG. For comparison, using $\alpha=0.47$, in that reference
{\v Z}itko found a phase transition between $2<S<5/2$,  whereas in our case we find  the transition for $S_c\sim0.28$ (see blue line in Fig. \ref{Fig:Shiba}). This puts in evidence the limitations of our method, which becomes unreliable near the QPT (i.e., in this case, when $S\approx S_c$). We recall here that the actual transition  arises from the competition between the Kondo correlations, which are absent in our approach,  and $\Delta$, and therefore is not related to the vanishing of Eq. (\ref{EQ_DeqZero}). In order to ensure the validity of our approach, we must restrict the value of $S$ to the regime $S\gg S_c$. As can be seen in Fig. \ref{Fig:Shiba}, the range of values of $S$ for which this requirement is fulfilled becomes parametrically larger when $\alpha \rightarrow 0$ (see black line, which correspond to the value $\alpha=0.1$, for which already the case of the lowest physical spin $S=1/2$ is already much larger than $S_c$).


\subsection{Regime $D \gg T$}

When the anisotropy is larger than the temperature, we need to distinguish the easy-axis ($D>0$) from the hard-axis case ($D<0$): while easy-axis anisotropy  tends to favour the classical limit when $D\rightarrow \infty$, this is not the case for hard-axis anisotropy, where strictly speaking the classical limit for the model Eq. (\ref{eq:H_total-1}) does not exist. Moreover, in this last case we still need to distinguish between integer and half-integer spin $S$, as the ground states are qualitatively different (non-degenerate or doubly degenerate, respectively). Therefore, in what follows we analyze three qualitatively different situations: 1) the easy-axis case $D>0$, 2) the hard-axis case $D<0$ with  half-integer $S$, and finally 3)  the hard-axis case $D<0$ with integer $S$.

\begin{figure}[t]
\includegraphics[width=1.0\columnwidth]{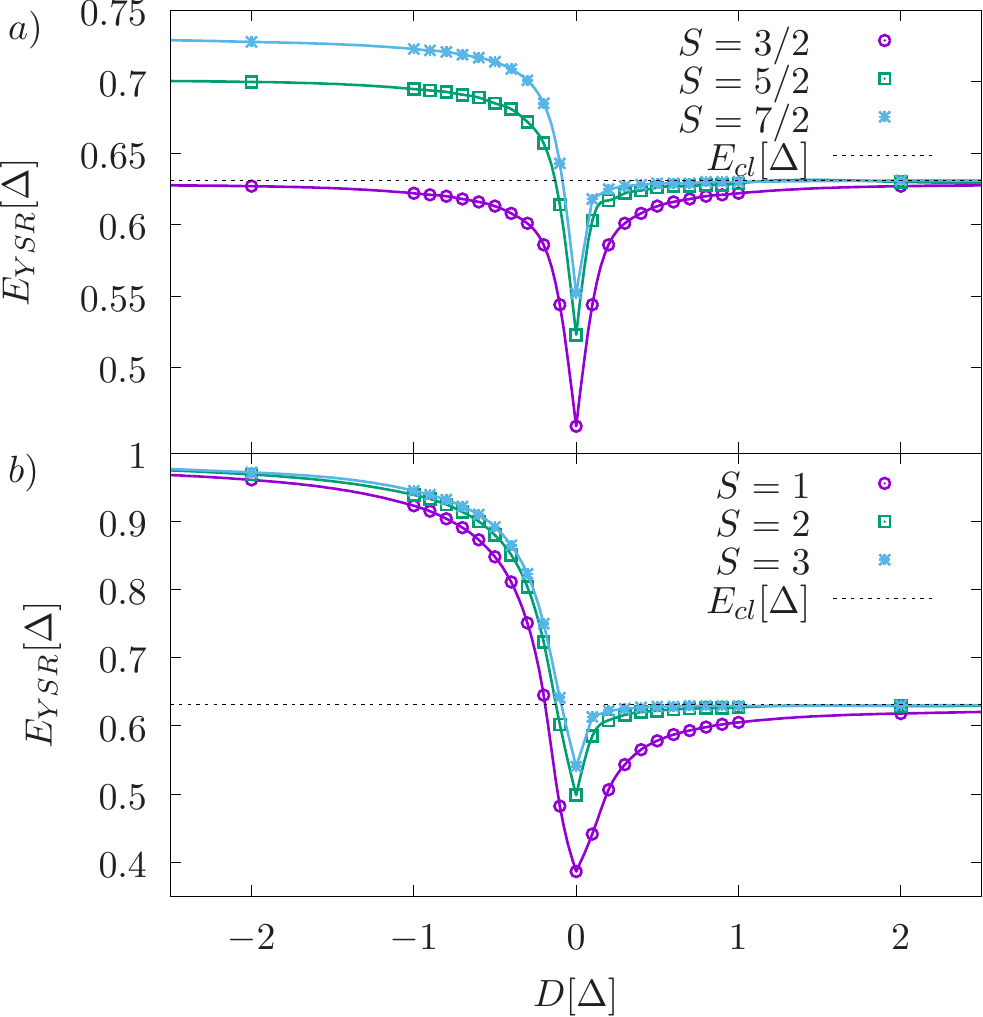}
\caption{(Color online) YSR subgap-state energy $E_\text{YSR}$ as a function of the anisotropy parameter $D$ for a) $S$ half-integer and b) $S$ integer. Here has been used the value $\alpha=0.5$.}
\label{Fig:Shiba_DyS}
\end{figure}

\subsubsection{Easy-axis case $D>0$}\label{Sub1}
The results in this section are depicted in Fig. \ref{Fig:Shiba_DyS} for $D>0$, both for half-integer $S$ [Fig. \ref{Fig:Shiba_DyS}(a)] and integer $S$ [Fig. \ref{Fig:Shiba_DyS}(b)]. In this case, the ground state of the isolated impurity takes the maximal $S^z$  projection $m=\pm S$, and therefore when either $S\rightarrow\infty$ or $D\rightarrow\infty$, the Shiba state energy converges to the same classical limit. However, when 
$D$ is larger than the pairing term $\Delta$, spin-fluctuations becomes negligible because the magnetic impurity needs an 
energy of the order of $\bar{\Delta}\equiv \Delta^0_{S}-\Delta^0_{S-1}=D(2S-1)$  to be able to flip to the first-excited states $m=\pm(S-1)$ [see Fig. \ref{Fig:Diagram}(b)]. 
Therefore, it is intuitively clear that the position of the Shiba state will converge faster to the classical limit $S\rightarrow\infty$ when $D>0$, as compared to the isotropic case in the previous section (Sec \ref{subsec:DllT}). 
This behavior can be clearly seen in Fig.  \ref{Fig:Shiba_DyS}, and can also be  understood directly from Eq. (\ref{eq:eq_integral}), where  the spin-flip term 
\begin{align}
\frac{\alpha_{\perp}^2}{ S^2}
\sum_{m=-S}^{S}\ \frac{A_{m}}{i\omega_l-i\omega_k-(\Delta_{m+1}^0-\Delta_{m}^0)}&\nonumber \\
\xrightarrow[T\rightarrow 0]{}\frac{\alpha_{\perp}^2}{S}\frac{\bar{\Delta}}{(\omega_l-\omega_k)^2+\bar{\Delta}^2}&,\label{eq:DMZ}
\end{align}
tends to zero when either $S\rightarrow \infty$ or $D\rightarrow \infty$ due to the presence of  $\bar{\Delta}$ is in the denominator. 

On the other hand, as discussed in the previous section, quantum fluctuations are enhanced at low temperatures in the absence of anisotropy, and this has important consequences for the position of the Shiba state, as can be seen in the sizable deviations from the classical limit when $D\rightarrow 0$ in Fig. \ref{Fig:Shiba_DyS}.

\begin{figure}[t]
\includegraphics[width=1\columnwidth]{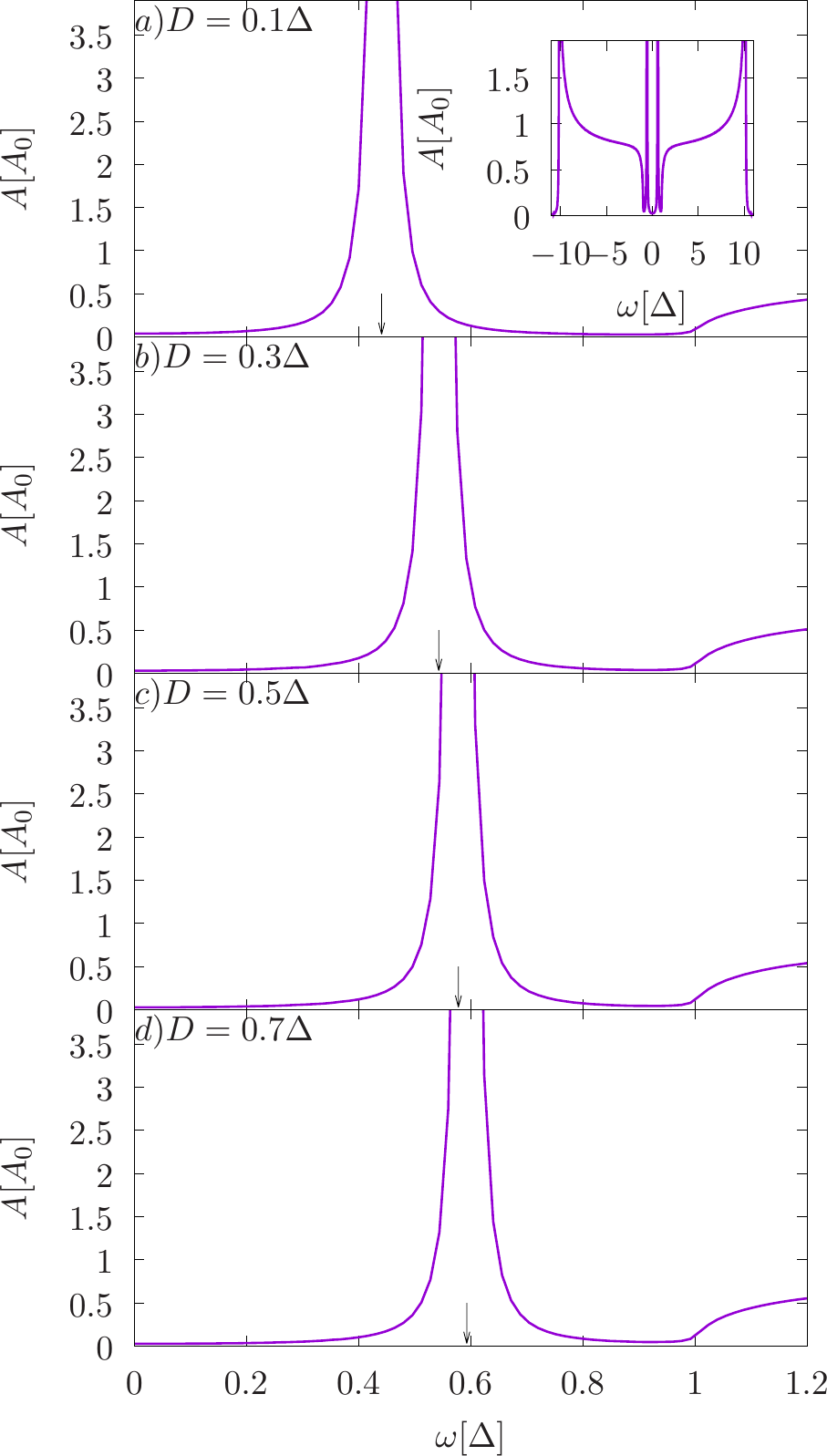}
\caption{(Color online) Local density of states (LDOS) at the site of the impurity for a) $D=0.1\Delta$, b) $D=0.3\Delta$, c) $D=0.5\Delta$ and d) $D=0.7\Delta$, at temperature $T=0.05\Delta$. $\alpha=0.5$ and $S=1$. The arrows indicate the center of the YSR resonance. Inset: full spectral density for $D=0.1\Delta$.}
\label{Fig:Dens}
\end{figure}

In order to illustrate the experimental consequences of our findings, in Fig. (\ref{Fig:Dens}) we show the local density of states at the impurity site for the specific case $S=1$. To minimize the number of  parameters in the model, we have chosen $\alpha_\parallel=\alpha_\perp=\alpha=0.5$ in order to show the effect of different values of the anisotropy parameter (here, $D=0.1\Delta$, $0.3\Delta$, $0.5\Delta$ and $0.7\Delta$). An artificial broadening $\epsilon=0.01 \Delta$  in Eq. (\ref{eq:rho_tilde}) has been used for visualization purposes, since the YSR state appears as a delta-peak in the local density of states when $\epsilon \rightarrow 0$. The center of the resonance in the figures is indicated with an arrow, and corresponds to the position of the YSR state, which shifts as a function of $D$.

This result is consistent with Ref. \onlinecite{Zitko_2017}, where the main conclusion is that realistic systems of anisotropic magnetic adsorbates tend to be well described  by the classical approximation, not because of their putatively ``large'' spin, but rather because of the energy-barrier effects induced by the magnetic anisotropy. A good example is Fe deposited on Cu$_2$N/Cu(100) \cite{Hirjibehedin07_Magnetic_anisotropy_in_surfaces, Zitko10_IETS_for_Fe_atoms}, where the classical picture of a static point-like magnetic field might be appropriate, but the value of the spin $S=2$ of the Fe atom is still far from the limit where the spin-flip scattering is negligible for the isotropic case (see Fig. \ref{Fig:Shiba}).

\subsubsection{Hard-axis anistropy $D<0$ and $S$  half-integer}\label{Sub2}

As mentioned before, hard-axis anisotropy favors states with minimal projection of $S^z$. In addition, when the spin $S$ is half-integer, the isolated impurity has a doubly-degenerate ground-state spanned by the states $m=\pm 1/2$, which implies that quantum fluctuations  subsist even in the limit $|D| \rightarrow \infty$. This means that  the classical YSR energy is never achieved in that limit [see the negative axis in Fig. \ref{Fig:Shiba_DyS}(a)]. 
Physically, taking the limit $D\rightarrow -\infty$ amounts to projecting out the states $|m|>1/2$: the fluctuations between states $\pm1/2\leftrightarrows\pm3/2\leftrightarrows \text{etc.}$ become negligible, but the fluctuations between states $-1/2\leftrightarrows 1/2$ remain. Eq. (\ref{eq:eq_integral}) writes
in this case:
\begin{align}
\tilde{g}_{\pm}\left(z\right) & =\tilde{g}_{\pm}^{0}\left(z\right)+\frac{4}{\pi^2}\left(\tilde{g}_{\pm}^{0}\left(z\right)\right)^{2}\left[\frac{\alpha_{\parallel}^{2}}{4S^2}\right.\nonumber\\
&\left.+\frac{\alpha_{\perp}^{2}}{S^{2}}\frac{S(S+1)+1/4}{2}\right]\tilde{g}_{\pm}\left(z\right),
\end{align}
and the original spin $S$ can be mapped onto an effective $S_\text{eff}=1/2$ impurity with renormalized couplings $\tilde{\alpha}_\parallel=\alpha_\parallel/\left(2S\right)$ and $\tilde{\alpha}_\perp=\alpha_\perp\sqrt{S\left(S+1\right)+1/4}/\left(2S\right)$. This result clearly illustrates that the classical-spin limit cannot be recovered taking $S\rightarrow \infty$ when $D<0$, as the transverse contribution becomes relatively more important than the parallel (classical) one. Consistently, when $W\rightarrow \infty$ and $\alpha_{\parallel}=\alpha_{\perp}=\alpha$, the position of the YSR states as a function of $S$ becomes
\begin{eqnarray}
\frac{E_\text{YSR}}{\Delta}&=&\frac{1-\frac{\alpha^2}{2}\left(1+\frac{1}{S}+\frac{3}{4S^2}\right)}{1+\frac{\alpha^2}{2}\left(1+\frac{1}{S}+\frac{3}{4S^2}\right)},\label{limit2}
\end{eqnarray}
very different from the classical limit $E_\text{cl}$ given by Eq. (\ref{ShibaLimit}) (note that the case $S=3/2$ is an exception, for which accidentally $E_\text{YSR}=E_\text{cl}$). This behavior is shown in Fig. \ref{Fig:Shiba_DyS}(a) for $D<0$. Note the stark contrast with respect to the case $D>0$.

An experimental example of this case (albeit in absence of superconductivity) is Co on Cu$_2$N/Cu(100), where the Co spin is $S=3/2$ but effectively it behaves as $S_\text{eff}=1/2$\cite{Otte08_The_role_of_anisotropy_on_Kondo_effect}.
	
\subsubsection{Hard-axis anisotropy $D<0$ and $S$  integer}\label{Sub3}
In Fig. \ref{Fig:Shiba_DyS}(b) we show the YSR energy for $D<0$ and spin $S$ integer. In this case the ground state of the isolated impurity corresponds to $m=0$, and therefore the impurity spin effectively becomes $S_\text{eff}=0$, and eventually decouples from the superconductor. Only the dynamical term proportional to $\alpha_{\perp}$ contributes to Eq. (\ref{eq:eq_integral}), and the system needs an energy of the order of $|D|$ in order to flip the spin to the states $m=\pm1$ at $T=0$. Therefore, when $D\rightarrow-\infty$ these fluctuations are forbidden, and the effective coupling between the superconductor and the impurity tends to zero. Consistently, the YSR levels shifts towards the edge of superconductor gap: the system effectively behaves as an unperturbed superconductor.

\subsection{Intermediate regime $D \simeq T$}\label{sec:results_T_finite}
\begin{figure}[t]g
\includegraphics[width=1\columnwidth]{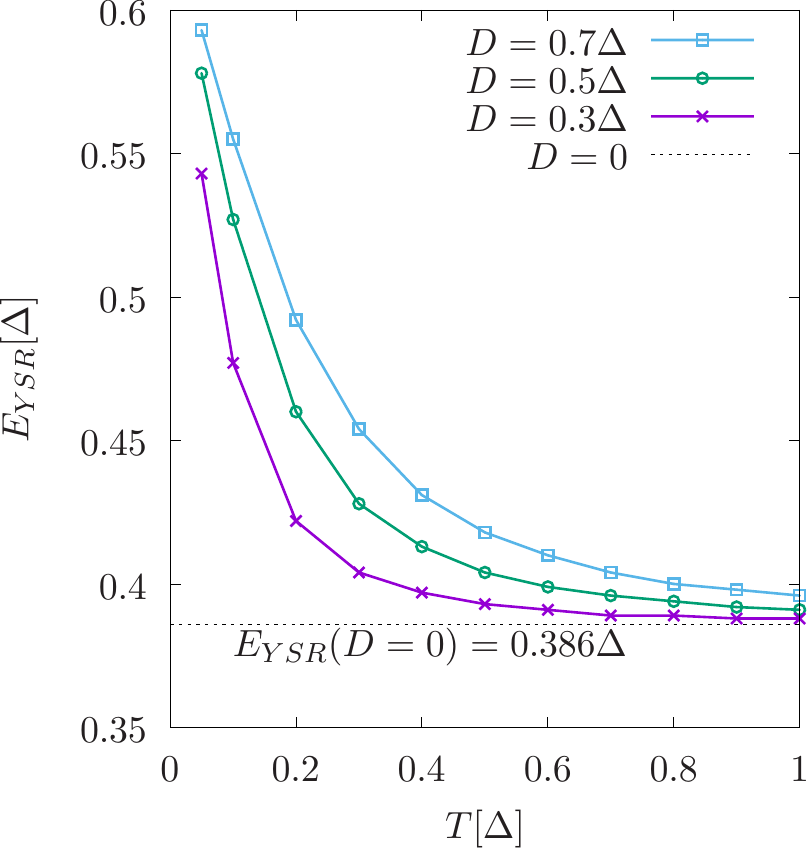}
\caption{(Color online) YSR subgap-state energy $E_\text{YSR}$ as a function of the temperature $T$, computed for $S=1$ and $\alpha=0.5$.}
\label{Fig:Shiba_T}
\end{figure}
Finally, we analyze the solutions of Eq. (\ref{eq:eq_integral}) obtained in the finite-temperature regime $T\simeq D$. Our main results are summarized in Fig. {\ref{Fig:Shiba_T}}, where we show a crossover of the Shiba peak from the regime $T\ll D$ to the regime $D\ll T$. This behavior arises from the interplay between anisotropy, quantum fluctuations and thermal effects. As can be seen in Eq. (\ref{eq:eq_integral}), the temperature dependence of the electronic Green's function $\tilde{g}_\pm\left(z\right)$ arises both from the temperature dependence of the thermodynamical average $\langle \left(S^{z}\right)^{2}\rangle_0$ [see Eq. (\ref{sz2})], as well as from the sum 
over the bosonic Matsubara  frequencies $i\omega_l$ of the dynamical correlators $\llangle S^{-};S^{+}\rrangle_0 \left(i\omega_{l}\right)$  [see Eq. (\ref{eq:smsp})]. 

The YSR energies shown in Fig. \ref{Fig:Shiba_T} have been computed for different values of the anisotropy $D=0$, $0.3\Delta$, $0.5\Delta$ and $0.7\Delta$, and for the specific case $S=1$ and $\alpha=0.5$. Note that the existence of anisotropy is a necessary condition to observe the shift: the constant behavior is recovered in the limit $D/T \rightarrow 0$. This can  be intuitively understood, since an excited spin state lying at an energy $\Delta^0_m$  will be thermally populated with Boltzmann probability $\sim e^{-\Delta^0_m/T}$. Therefore, new scattering channels, which are forbidden at $T=0$, will be allowed when $T\simeq \Delta_m^0$. Increasing the temperature beyond $T>\text{max}\left\{ \Delta^0_m\right\}$ implies that the system has enough energy to explore the whole spin multiplet, and the SU(2) symmetry is effectively restored. Then, the position of the YSR state becomes described by Eq. (\ref{EQ_DeqZero}) in the isotropic limit $D\rightarrow 0$, and converges to the value $E_{YSR}=0.386\Delta$.  On the other hand, when  $T\ll \Delta^0_m$, the thermal population of the excited states  become exponentially small, and therefore they become effectively inaccessible.  Thus, the system reaches its the classical limit and the YSR energy tends to the classical result  $E_{YSR}=0.631\Delta$ (see dashed lines in Fig. \ref{Fig:Shiba_DyS}). In the limit of infinite bandwidth the result is given by  Eq. (\ref{ShibaLimit}) and yields  $E_{YSR}=0.6\Delta$.

This crossover is a feature which could be experimentally tested in STM experiments, and which has not been discussed before. In the case of the single-orbital impurity Anderson model (which maps onto the $S=1/2$ Kondo impurity)\cite{Zitko16_Spectral_properties_of_Shiba_states_at_finite_T}, the single-ion anisotropy has no effect, and therefore no shift is observed. In addition, in the classical spin approximation, the effective a one-body description leads naturally to a temperature-independent energy spectrum \cite{Yu65_YSR_states, Shiba_1968, Rusinov69_YSR_states}. Then, these previous results might give the impression that the position of the Shiba state does not depend on temperature\cite{Hatter2017_Scaling_of_YSR_energies}. However, based on our theoretical considerations, we claim that this is not true in the more general case  of anisotropic quantum impurities with $S>1/2$. Moreover, we speculate that this effect  might have already been seen in the STM experiments of Ref. 
\onlinecite{Hatter2017_Scaling_of_YSR_energies} (see Fig. 3.d in that reference), where the authors observed that thermal scattering diminishes the effect of the magnetocrystalline anisotropy.

Besides the shift of the Shiba peaks described in Fig. (\ref{Fig:Shiba_T}) we do not observe any other qualitative change in the density of states (e.g., thermal broadening of the resonances) as compared to the regimes $D\ll T$ or $T\ll D$. We believe that this a consequence of our approximation  (\ref{approx_spin_correlators}), which might be too simplistic to properly describe thermal effects in more detail. This last conjecture is based on recent NRG results obtained in the limit $D=0$\cite{Zitko16_Spectral_properties_of_Shiba_states_at_finite_T}, where the emergence  of an intragap thermal continuum with spectral weight $\sim e^{-\Delta/T}$ around the Shiba peaks has been reported.

\section{Summary and conclusions}\label{sec:summary}

We have studied the effect of uniaxial magnetic anisotropy and quantum fluctuations on the Yu-Shiba-Rusinov states emerging in systems of magnetic impurities coupled to superconductors. YSR states have become an active area of  research in solid-state physics due to their potential applications in the study of topological phases and in future quantum-information technologies. However, many theoretical aspects concerning the effects of anistropy, temperature and quantum fluctuations are yet to be understood for the correct interpretation of experiments. In particular, due to its ever-present nature at the surface of metals with large  atomic number (such as Pb),  single-ion anisotropy is an important effect that cannot be disregarded in realistic theoretical descriptions of YSR states.

In this work we have proposed and implemented a novel decoupling scheme for the equations of motion of the conduction-electron's Green's function, valid in the weak coupling regime $T_K \ll \Delta $ where the impurity spin is unscreened. This decoupling allows to go beyond the classical spin approximation and to obtain the Green's function (and the position of the YSR states) in the presence of single-ion anisotropy and quantum fluctuations. For realistic systems, this implies that our results should be applicable in the unscreened limit of, e.g.,  the experiments reported in Refs. \onlinecite{Franke_2011, Bauer13_Kondo_screening_and_pairing_on_Mn_phtalocyanines_on_Pb, Hatter_2015,Hatter2017_Scaling_of_YSR_energies}. In these works the authors study MnPc organometallic molecules which are weakly coupled to superconducting Pb surfaces, and therefore could be suitable physical realizations of the systems studied in this work.

In the idealistic case of an impurity with vanishing anisotropy ($D=0$), the classical limit is slowly reached as $\sim1/S$ (Fig. \ref{Fig:Shiba}), and we expect that quantum fluctuations remain important even for impurities with a relatively ``large'' spin. Although the complete absence  of anisotropy is  an unrealistic experimental situation, this case is instructive as it enables an important conclusion: the classical-spin behavior observed in certain experiments\cite{Hirjibehedin07_Magnetic_anisotropy_in_surfaces} is not originated in the size of $S$, but rather is a consequence of the magnitude of the anisotropy parameter $D$. This conclusion fully agrees with recent NRG calculations \cite{Zitko16_Spectral_properties_of_Shiba_states_at_finite_T}, and constitutes an important validity check of our approximation in the case $D=0$.

Our main conclusion is that in the absence of other quantum effects (such as the aforementioned Kondo effect), the YSR states will be strongly affected both by the anisotropy and the temperature. This prediction is in contrast to the common belief that YSR states are unaffected by the temperature.  In the case of easy-axis anisotropy $D>0$, in the limit $T\ll D$ our results converge fast to the classical-spin case [see Fig. \ref{Fig:Shiba_DyS} and Eq. (\ref{eq:DMZ})]. The fact that our calculations recover the classical limit is an important sanity check. On the other hand, for $D\ll T$ where thermal energy effectively restores the SU(2) symmetry and the quantum fluctuations remain an important effect, the position of the YSR peak might deviate considerably from the classical expression. On the other hand, for the case of hard-axis anisotropy $D<0$, the nature of the quantum impurity radically changes and strictly speaking there is no classical limit. Indeed, in the limit $D\rightarrow -\infty$ the impurity effectively maps onto a $S_\text{eff}=1/2$ ($S_\text{eff}=0$) impurity for half-integer (integer) $S$\cite{Otte08_The_role_of_anisotropy_on_Kondo_effect}. Consequently the position of the YSR peak is not described by the classical formula (\ref{ShibaLimit}), but instead is described by Eq. (\ref{limit2}) or tends to $E_{YSR}/\Delta\rightarrow 1$, respectively. 

From these results, we conclude that the position of the YSR peaks is the result of a complicated interplay between quantum fluctuations, anisotropy and temperature, and that it is crucial to take all of these effects into account for the correct interpretation of the experimental STM data.  For instance, in Ref. \onlinecite{Hatter2017_Scaling_of_YSR_energies} the authors describe a  $S=1$ system with an anisotropy parameter $D=0.33$ meV and $\Delta=1.3$ meV, with temperatures ranging from 1 to 10 K (i.e., 0.09 meV to 0.9 meV). This implies that the impurity crosses over from the regime $T\simeq 0.27 D$ to the regime $T\simeq 2.7 D$. Quite remarkably, precisely for $T\simeq D$, the experimental results show a qualitative change of behavior and for $T>D$  thermal scattering seem to diminish the effect of anisotropy, a fact that seems to be consistent with our prediction in Fig. \ref{Fig:Shiba_T}.

We believe that the decoupling method outlined in this work could be relevant in the field of YSR states studied by STM techniques, where the inherent many-body nature of the problem and the unavoidable experimental complexities must be considered for the correct interpretation of the experimental results.

\acknowledgments
A.M.L. acknowledges financial support from PICT-2015-0217  and PICT-2017-2081 (ANPCyT - Argentina), PIP-11220150100364 (CONICET - Argentina) and Relocation Grant  RD1158 - 52368 (CONICET - Argentina). A.M.L. thanks Pablo Roura-Bas and Armando A. Aligia for useful discussions. This work used the Toko Cluster from FCEN-UNCuyo, which is part of the SNCAD-MinCyT, Argentina.

\appendix
\section{Calculation of the vector $\mathbf{S}_k$ in Eq. (\ref{matrix_equation_3})}\label{sec:App}

In this Appendix, we show the details for the computation of the Matsubara sum $\sum_{l_\text{max}<|l|}$ in Eq. (\ref{matrix_equation_3}):
\begin{align}
M_k^m&=\frac{1}{\beta}\sum_{l_\text{max}<|l|} \frac{1}{i\omega_{l}-i\omega_{k}-\left(\Delta^0_{m+1}-\Delta^0_{m}\right)}.\frac{1}{\left(z-i\omega_l\right)}\nonumber\\
&=\frac{\beta}{(2\pi )^2}\sum_{l_\text{max}<|l|} \frac{1}{l-k-\frac{\beta}{2\pi i}\left(\Delta^0_{m+1}-\Delta^0_{m}\right)}.\frac{1}{l-\frac{\beta}{2\pi i} z}.\label{eq:sum3}
\end{align}
Splitting the sum into $\sum_{l_\text{max}<|l|}=\sum^\infty_{l=l_\text{max}+1}+\sum_{l=-\infty}^{-l_\text{max}-1}$, and performing the change of variable $l=l^\prime+l_\text{max}$, we can write
\begin{widetext}
\begin{align}
M_k^m=\frac{\beta}{(2\pi )^2}\sum^\infty_{l^\prime=1}& \left[\frac{1}{l^\prime+l_\text{max}-k-\frac{\beta}{2\pi i}\left(\Delta^0_{m+1}-\Delta^0_{m}\right)}.\frac{1}{l^\prime+l_\text{max}-\frac{\beta}{2\pi i} z}\right.\nonumber\\
&\left.+\frac{1}{l^\prime+l_\text{max}+k+\frac{\beta}{2\pi i}\left(\Delta^0_{m+1}-\Delta^0_{m}\right)}.\frac{1}{l^\prime+l_\text{max}+\frac{\beta}{2\pi i} z}\right].\label{eq:sum3}
\end{align}
The summation has the form
\begin{align}
\sum_{l^\prime=1}^\infty \left[\frac{1}{\left(l^\prime+z_{1,k}^{\left(m\right)}\right)\left(l^\prime+z_{2,k}^{\left(m\right)}\right)}+\frac{1}{\left(l^\prime+z_{3,k}^{\left(m\right)}\right)\left(l^\prime+z_{4,k}^{\left(m\right)}\right)}\right],\label{eq:sum1}
\end{align}
where
\begin{align}
z_{1,k}^{\left(m\right)}&=l_\text{max}-k-\frac{\beta \left(\Delta^0_{m+1}-\Delta^0_{m}\right)}{2\pi i},\\
z_{2,k}^{\left(m\right)}&=l_\text{max}-\frac{\beta z}{2\pi i},\\ 
z_{3,k}^{\left(m\right)}&=l_\text{max}+k+\frac{\beta \left(\Delta^0_{m+1}-\Delta^0_{m}\right)}{2\pi i},\\
z_{4,k}^{\left(m\right)}&=l_\text{max}+\frac{\beta z}{2\pi i}, 
\end{align}
Using the result\cite{abramowitz} $\Psi(1+z)=-\gamma+\sum_{l^\prime=1}^\infty\frac{z}{l^\prime \left(l^\prime+z\right)}$, where $\psi\left(z\right)$ and $\gamma$ are, respectively, the digamma function and the Euler gamma constant (see Ref. \onlinecite{abramowitz}), the sum (\ref{eq:sum1}) yields
\begin{align}
\sum_{l^\prime=1}^\infty \left[\frac{1}{\left(l^\prime+z_{1,k}^{\left(m\right)}\right)\left(l^\prime+z_{2,k}^{\left(m\right)}\right)}+\frac{1}{\left(l^\prime+z_{3,k}^{\left(m\right)}\right)\left(l^\prime+z_{4,k}^{\left(m\right)}\right)}\right]&=\frac{\Psi(1+z_{1,k}^{\left(m\right)})-\Psi(1+z_{2,k}^{\left(m\right)})}{z_{1,k}^{\left(m\right)}-z_{2,k}^{\left(m\right)}}+ \frac{\Psi(1+z_{3,k}^{\left(m\right)})-\Psi(1+z_{4,k}^{\left(m\right)})}{z_{3,k}^{\left(m\right)}-z_{4,k}^{\left(m\right)}}.\label{eq:sum2}
\end{align}
With all these results, finally the  Eq. (\ref{matrix_equation_3}) is

\begin{align}
\mathbf{S}_k&= \frac{4\alpha_{\perp}^{2}\left(\mathbf{g}_{k}^{\left(0\right)} \right)^{2}}{\pi^2S^2} \sum_{m=-S}^{S}  \frac{A_m W}{(2\pi)^2}\left(\frac{\Psi(1+z_{1,k}^{\left(m\right)})-\Psi(1+z_{2,k}^{\left(m\right)})}{\left(z_{1,k}^{\left(m\right)}-z_{2,k}^{\left(m\right)}\right)/\beta}+ \frac{\Psi(1+z_{3,k}^{\left(m\right)})-\Psi(1+z_{4,k}^{\left(m\right)})}{\left(z_{3,k}^{\left(m\right)}-z_{4,k}^{\left(m\right) }\right)/\beta}\right).
\label{final_Sk}\end{align}

\end{widetext}

\bibliographystyle{apsrev4-1}

\begin{thebibliography}{57}%
\makeatletter
\providecommand \@ifxundefined [1]{%
 \@ifx{#1\undefined}
}%
\providecommand \@ifnum [1]{%
 \ifnum #1\expandafter \@firstoftwo
 \else \expandafter \@secondoftwo
 \fi
}%
\providecommand \@ifx [1]{%
 \ifx #1\expandafter \@firstoftwo
 \else \expandafter \@secondoftwo
 \fi
}%
\providecommand \natexlab [1]{#1}%
\providecommand \enquote  [1]{``#1''}%
\providecommand \bibnamefont  [1]{#1}%
\providecommand \bibfnamefont [1]{#1}%
\providecommand \citenamefont [1]{#1}%
\providecommand \href@noop [0]{\@secondoftwo}%
\providecommand \href [0]{\begingroup \@sanitize@url \@href}%
\providecommand \@href[1]{\@@startlink{#1}\@@href}%
\providecommand \@@href[1]{\endgroup#1\@@endlink}%
\providecommand \@sanitize@url [0]{\catcode `\\12\catcode `\$12\catcode
  `\&12\catcode `\#12\catcode `\^12\catcode `\_12\catcode `\%12\relax}%
\providecommand \@@startlink[1]{}%
\providecommand \@@endlink[0]{}%
\providecommand \url  [0]{\begingroup\@sanitize@url \@url }%
\providecommand \@url [1]{\endgroup\@href {#1}{\urlprefix }}%
\providecommand \urlprefix  [0]{URL }%
\providecommand \Eprint [0]{\href }%
\providecommand \doibase [0]{http://dx.doi.org/}%
\providecommand \selectlanguage [0]{\@gobble}%
\providecommand \bibinfo  [0]{\@secondoftwo}%
\providecommand \bibfield  [0]{\@secondoftwo}%
\providecommand \translation [1]{[#1]}%
\providecommand \BibitemOpen [0]{}%
\providecommand \bibitemStop [0]{}%
\providecommand \bibitemNoStop [0]{.\EOS\space}%
\providecommand \EOS [0]{\spacefactor3000\relax}%
\providecommand \BibitemShut  [1]{\csname bibitem#1\endcsname}%
\let\auto@bib@innerbib\@empty
\bibitem [{\citenamefont {Balatsky}\ \emph {et~al.}(2006)\citenamefont
  {Balatsky}, \citenamefont {Vekhter},\ and\ \citenamefont
  {Zhu}}]{Balatsky_2006}%
  \BibitemOpen
  \bibfield  {author} {\bibinfo {author} {\bibfnamefont {A.~V.}\ \bibnamefont
  {Balatsky}}, \bibinfo {author} {\bibfnamefont {I.}~\bibnamefont {Vekhter}}, \
  and\ \bibinfo {author} {\bibfnamefont {J.-X.}\ \bibnamefont {Zhu}},\ }\href
  {\doibase 10.1103/RevModPhys.78.373} {\bibfield  {journal} {\bibinfo
  {journal} {Rev. Mod. Phys.}\ }\textbf {\bibinfo {volume} {78}},\ \bibinfo
  {pages} {373} (\bibinfo {year} {2006})}\BibitemShut {NoStop}%
\bibitem [{\citenamefont {Heinrich}\ \emph {et~al.}(2018)\citenamefont
  {Heinrich}, \citenamefont {Pascual},\ and\ \citenamefont
  {Franke}}]{Heinrich18_Review_single_adsorbates}%
  \BibitemOpen
  \bibfield  {author} {\bibinfo {author} {\bibfnamefont {B.~W.}\ \bibnamefont
  {Heinrich}}, \bibinfo {author} {\bibfnamefont {J.~I.}\ \bibnamefont
  {Pascual}}, \ and\ \bibinfo {author} {\bibfnamefont {K.~J.}\ \bibnamefont
  {Franke}},\ }\href {\doibase https://doi.org/10.1016/j.progsurf.2018.01.001}
  {\bibfield  {journal} {\bibinfo  {journal} {Progress in Surface Science}\
  }\textbf {\bibinfo {volume} {93}},\ \bibinfo {pages} {1 } (\bibinfo {year}
  {2018})}\BibitemShut {NoStop}%
\bibitem [{\citenamefont {Yu}(1965)}]{Yu65_YSR_states}%
  \BibitemOpen
  \bibfield  {author} {\bibinfo {author} {\bibfnamefont {L.}~\bibnamefont
  {Yu}},\ }\href@noop {} {\bibfield  {journal} {\bibinfo  {journal} {Acta Phys.
  Sin.}\ }\textbf {\bibinfo {volume} {21}},\ \bibinfo {pages} {75} (\bibinfo
  {year} {1965})}\BibitemShut {NoStop}%
\bibitem [{\citenamefont {Shiba}(1968)}]{Shiba_1968}%
  \BibitemOpen
  \bibfield  {author} {\bibinfo {author} {\bibfnamefont {H.}~\bibnamefont
  {Shiba}},\ }\href {\doibase 10.1143/PTP.40.435} {\bibfield  {journal}
  {\bibinfo  {journal} {Progress of Theoretical Physics}\ }\textbf {\bibinfo
  {volume} {40}},\ \bibinfo {pages} {435} (\bibinfo {year} {1968})}\BibitemShut
  {NoStop}%
\bibitem [{\citenamefont {Rusinov}(1969)}]{Rusinov69_YSR_states}%
  \BibitemOpen
  \bibfield  {author} {\bibinfo {author} {\bibfnamefont {A.~I.}\ \bibnamefont
  {Rusinov}},\ }\href@noop {} {\bibfield  {journal} {\bibinfo  {journal} {Zh.
  Eksp. Teor. Fiz.}\ }\textbf {\bibinfo {volume} {56}},\ \bibinfo {pages}
  {2047} (\bibinfo {year} {1969})}\BibitemShut {NoStop}%
\bibitem [{\citenamefont {Yazdani}\ \emph {et~al.}(1997)\citenamefont
  {Yazdani}, \citenamefont {Jones}, \citenamefont {Lutz}, \citenamefont
  {Crommie},\ and\ \citenamefont {Eigler}}]{Yazdani97_YSR_states}%
  \BibitemOpen
  \bibfield  {author} {\bibinfo {author} {\bibfnamefont {A.}~\bibnamefont
  {Yazdani}}, \bibinfo {author} {\bibfnamefont {B.~A.}\ \bibnamefont {Jones}},
  \bibinfo {author} {\bibfnamefont {C.~P.}\ \bibnamefont {Lutz}}, \bibinfo
  {author} {\bibfnamefont {M.~F.}\ \bibnamefont {Crommie}}, \ and\ \bibinfo
  {author} {\bibfnamefont {D.~M.}\ \bibnamefont {Eigler}},\ }\href
  {http://science.sciencemag.org/content/275/5307/1767.full} {\bibfield
  {journal} {\bibinfo  {journal} {Science}\ }\textbf {\bibinfo {volume}
  {275}},\ \bibinfo {pages} {1767} (\bibinfo {year} {1997})}\BibitemShut
  {NoStop}%
\bibitem [{\citenamefont {Ji}\ \emph {et~al.}(2008)\citenamefont {Ji},
  \citenamefont {Zhang}, \citenamefont {Fu}, \citenamefont {Chen},
  \citenamefont {Ma}, \citenamefont {Li}, \citenamefont {Duan}, \citenamefont
  {Jia},\ and\ \citenamefont {Xue}}]{Ji08_YSR_states}%
  \BibitemOpen
  \bibfield  {author} {\bibinfo {author} {\bibfnamefont {S.-H.}\ \bibnamefont
  {Ji}}, \bibinfo {author} {\bibfnamefont {T.}~\bibnamefont {Zhang}}, \bibinfo
  {author} {\bibfnamefont {Y.-S.}\ \bibnamefont {Fu}}, \bibinfo {author}
  {\bibfnamefont {X.}~\bibnamefont {Chen}}, \bibinfo {author} {\bibfnamefont
  {X.-C.}\ \bibnamefont {Ma}}, \bibinfo {author} {\bibfnamefont
  {J.}~\bibnamefont {Li}}, \bibinfo {author} {\bibfnamefont {W.-H.}\
  \bibnamefont {Duan}}, \bibinfo {author} {\bibfnamefont {J.-F.}\ \bibnamefont
  {Jia}}, \ and\ \bibinfo {author} {\bibfnamefont {Q.-K.}\ \bibnamefont
  {Xue}},\ }\href {\doibase 10.1103/PhysRevLett.100.226801} {\bibfield
  {journal} {\bibinfo  {journal} {Phys. Rev. Lett.}\ }\textbf {\bibinfo
  {volume} {100}},\ \bibinfo {pages} {226801} (\bibinfo {year}
  {2008})}\BibitemShut {NoStop}%
\bibitem [{\citenamefont {Iavarone}\ \emph {et~al.}(2010)\citenamefont
  {Iavarone}, \citenamefont {Karapetrov}, \citenamefont {Fedor}, \citenamefont
  {Rosenmann}, \citenamefont {Nishizaki},\ and\ \citenamefont
  {Kobayashi}}]{Iavarone10_Local_effects_of_magnetic_impurities_on_SCs}%
  \BibitemOpen
  \bibfield  {author} {\bibinfo {author} {\bibfnamefont {M.}~\bibnamefont
  {Iavarone}}, \bibinfo {author} {\bibfnamefont {G.}~\bibnamefont
  {Karapetrov}}, \bibinfo {author} {\bibfnamefont {J.}~\bibnamefont {Fedor}},
  \bibinfo {author} {\bibfnamefont {D.}~\bibnamefont {Rosenmann}}, \bibinfo
  {author} {\bibfnamefont {T.}~\bibnamefont {Nishizaki}}, \ and\ \bibinfo
  {author} {\bibfnamefont {N.}~\bibnamefont {Kobayashi}},\ }\href {\doibase
  10.1088/0953-8984/22/1/015501} {\bibfield  {journal} {\bibinfo  {journal} {J.
  Phys.: Condens. Matter}\ }\textbf {\bibinfo {volume} {22}},\ \bibinfo {pages}
  {015501} (\bibinfo {year} {2010})}\BibitemShut {NoStop}%
\bibitem [{\citenamefont {Ji}\ \emph {et~al.}(2010)\citenamefont {Ji},
  \citenamefont {Zhang}, \citenamefont {Fu}, \citenamefont {Chen},
  \citenamefont {Jia}, \citenamefont {Xue},\ and\ \citenamefont
  {Ma}}]{Ji10_YSR_states_for_the_chemical_identification_of_adatoms}%
  \BibitemOpen
  \bibfield  {author} {\bibinfo {author} {\bibfnamefont {S.-H.}\ \bibnamefont
  {Ji}}, \bibinfo {author} {\bibfnamefont {T.}~\bibnamefont {Zhang}}, \bibinfo
  {author} {\bibfnamefont {Y.-S.}\ \bibnamefont {Fu}}, \bibinfo {author}
  {\bibfnamefont {X.}~\bibnamefont {Chen}}, \bibinfo {author} {\bibfnamefont
  {J.-F.}\ \bibnamefont {Jia}}, \bibinfo {author} {\bibfnamefont {Q.-K.}\
  \bibnamefont {Xue}}, \ and\ \bibinfo {author} {\bibfnamefont {X.-C.}\
  \bibnamefont {Ma}},\ }\href {\doibase 10.1063/1.3318404} {\bibfield
  {journal} {\bibinfo  {journal} {App. Phys. Lett.}\ }\textbf {\bibinfo
  {volume} {96}},\ \bibinfo {pages} {073113} (\bibinfo {year}
  {2010})}\BibitemShut {NoStop}%
\bibitem [{\citenamefont {Franke}\ \emph {et~al.}(2011)\citenamefont {Franke},
  \citenamefont {Schulze},\ and\ \citenamefont {Pascual}}]{Franke_2011}%
  \BibitemOpen
  \bibfield  {author} {\bibinfo {author} {\bibfnamefont {K.~J.}\ \bibnamefont
  {Franke}}, \bibinfo {author} {\bibfnamefont {G.}~\bibnamefont {Schulze}}, \
  and\ \bibinfo {author} {\bibfnamefont {J.~I.}\ \bibnamefont {Pascual}},\
  }\href {\doibase 10.1126/science.1202204} {\bibfield  {journal} {\bibinfo
  {journal} {Science}\ }\textbf {\bibinfo {volume} {332}},\ \bibinfo {pages}
  {940} (\bibinfo {year} {2011})}\BibitemShut {NoStop}%
\bibitem [{\citenamefont {Bauer}\ \emph {et~al.}(2013)\citenamefont {Bauer},
  \citenamefont {Pascual},\ and\ \citenamefont
  {Franke}}]{Bauer13_Kondo_screening_and_pairing_on_Mn_phtalocyanines_on_Pb}%
  \BibitemOpen
  \bibfield  {author} {\bibinfo {author} {\bibfnamefont {J.}~\bibnamefont
  {Bauer}}, \bibinfo {author} {\bibfnamefont {J.~I.}\ \bibnamefont {Pascual}},
  \ and\ \bibinfo {author} {\bibfnamefont {K.~J.}\ \bibnamefont {Franke}},\
  }\href {\doibase 10.1103/PhysRevB.87.075125} {\bibfield  {journal} {\bibinfo
  {journal} {Phys. Rev. B}\ }\textbf {\bibinfo {volume} {87}},\ \bibinfo
  {pages} {075125} (\bibinfo {year} {2013})}\BibitemShut {NoStop}%
\bibitem [{\citenamefont {Hatter}\ \emph {et~al.}(2015)\citenamefont {Hatter},
  \citenamefont {Heinrich}, \citenamefont {Ruby}, \citenamefont {Pascual},\
  and\ \citenamefont {Franke}}]{Hatter_2015}%
  \BibitemOpen
  \bibfield  {author} {\bibinfo {author} {\bibfnamefont {N.}~\bibnamefont
  {Hatter}}, \bibinfo {author} {\bibfnamefont {B.~W.}\ \bibnamefont
  {Heinrich}}, \bibinfo {author} {\bibfnamefont {M.}~\bibnamefont {Ruby}},
  \bibinfo {author} {\bibfnamefont {J.~I.}\ \bibnamefont {Pascual}}, \ and\
  \bibinfo {author} {\bibfnamefont {K.~J.}\ \bibnamefont {Franke}},\
  }\href@noop {} {\bibfield  {journal} {\bibinfo  {journal} {Nature
  Communications}\ }\textbf {\bibinfo {volume} {6}},\ \bibinfo {pages} {8988}
  (\bibinfo {year} {2015})}\BibitemShut {NoStop}%
\bibitem [{\citenamefont {Hatter}\ \emph {et~al.}(2017)\citenamefont {Hatter},
  \citenamefont {Heinrich}, \citenamefont {Rolf},\ and\ \citenamefont
  {Franke}}]{Hatter2017_Scaling_of_YSR_energies}%
  \BibitemOpen
  \bibfield  {author} {\bibinfo {author} {\bibfnamefont {N.}~\bibnamefont
  {Hatter}}, \bibinfo {author} {\bibfnamefont {B.~W.}\ \bibnamefont
  {Heinrich}}, \bibinfo {author} {\bibfnamefont {D.}~\bibnamefont {Rolf}}, \
  and\ \bibinfo {author} {\bibfnamefont {K.~J.}\ \bibnamefont {Franke}},\
  }\href {\doibase 10.1038/s41467-017-02277-7} {\bibfield  {journal} {\bibinfo
  {journal} {Nature Communications}\ }\textbf {\bibinfo {volume} {8}},\
  \bibinfo {pages} {2016} (\bibinfo {year} {2017})}\BibitemShut {NoStop}%
\bibitem [{\citenamefont {Ruby}\ \emph {et~al.}(2016)\citenamefont {Ruby},
  \citenamefont {Peng}, \citenamefont {von Oppen}, \citenamefont {Heinrich},\
  and\ \citenamefont {Franke}}]{Ruby_2016}%
  \BibitemOpen
  \bibfield  {author} {\bibinfo {author} {\bibfnamefont {M.}~\bibnamefont
  {Ruby}}, \bibinfo {author} {\bibfnamefont {Y.}~\bibnamefont {Peng}}, \bibinfo
  {author} {\bibfnamefont {F.}~\bibnamefont {von Oppen}}, \bibinfo {author}
  {\bibfnamefont {B.~W.}\ \bibnamefont {Heinrich}}, \ and\ \bibinfo {author}
  {\bibfnamefont {K.~J.}\ \bibnamefont {Franke}},\ }\href {\doibase
  10.1103/PhysRevLett.117.186801} {\bibfield  {journal} {\bibinfo  {journal}
  {Phys. Rev. Lett.}\ }\textbf {\bibinfo {volume} {117}},\ \bibinfo {pages}
  {186801} (\bibinfo {year} {2016})}\BibitemShut {NoStop}%
\bibitem [{\citenamefont {Choi}\ \emph {et~al.}(2017)\citenamefont {Choi},
  \citenamefont {Deung-Jang}, \citenamefont {Carmen}, \citenamefont {Joeri},
  \citenamefont {Miguel~M.}, \citenamefont {Nicolás},\ and\ \citenamefont
  {Ignacio}}]{Choi_2017}%
  \BibitemOpen
  \bibfield  {author} {\bibinfo {author} {\bibnamefont {Choi}}, \bibinfo
  {author} {\bibfnamefont {R.-V.}\ \bibnamefont {Deung-Jang}}, \bibinfo
  {author} {\bibfnamefont {d.~B.}\ \bibnamefont {Carmen}}, \bibinfo {author}
  {\bibfnamefont {U.}~\bibnamefont {Joeri}}, \bibinfo {author} {\bibfnamefont
  {L.}~\bibnamefont {Miguel~M.}}, \bibinfo {author} {\bibfnamefont
  {P.}~\bibnamefont {Nicolás}}, \ and\ \bibinfo {author} {\bibfnamefont
  {J.}~\bibnamefont {Ignacio}},\ }\href {\doibase 10.1038/ncomms15175}
  {\bibfield  {journal} {\bibinfo  {journal} {Nature Communications}\ }\textbf
  {\bibinfo {volume} {8}},\ \bibinfo {pages} {15175} (\bibinfo {year}
  {2017})}\BibitemShut {NoStop}%
\bibitem [{\citenamefont {Ruby}\ \emph
  {et~al.}(2015{\natexlab{a}})\citenamefont {Ruby}, \citenamefont {Pientka},
  \citenamefont {Peng}, \citenamefont {von Oppen}, \citenamefont {Heinrich},\
  and\ \citenamefont
  {Franke}}]{Ruby15_Tunneling_into_localized_subgap_states_in_SC}%
  \BibitemOpen
  \bibfield  {author} {\bibinfo {author} {\bibfnamefont {M.}~\bibnamefont
  {Ruby}}, \bibinfo {author} {\bibfnamefont {F.}~\bibnamefont {Pientka}},
  \bibinfo {author} {\bibfnamefont {Y.}~\bibnamefont {Peng}}, \bibinfo {author}
  {\bibfnamefont {F.}~\bibnamefont {von Oppen}}, \bibinfo {author}
  {\bibfnamefont {B.~W.}\ \bibnamefont {Heinrich}}, \ and\ \bibinfo {author}
  {\bibfnamefont {K.~J.}\ \bibnamefont {Franke}},\ }\href {\doibase
  10.1103/PhysRevLett.115.087001} {\bibfield  {journal} {\bibinfo  {journal}
  {Phys. Rev. Lett.}\ }\textbf {\bibinfo {volume} {115}},\ \bibinfo {pages}
  {087001} (\bibinfo {year} {2015}{\natexlab{a}})}\BibitemShut {NoStop}%
\bibitem [{\citenamefont {Heinrich}\ \emph {et~al.}(2013)\citenamefont
  {Heinrich}, \citenamefont {Braun}, \citenamefont {Pascual},\ and\
  \citenamefont
  {Franke}}]{Heinrich13_Protection_of_excited_spin_states_by_SC_gap}%
  \BibitemOpen
  \bibfield  {author} {\bibinfo {author} {\bibfnamefont {B.}~\bibnamefont
  {Heinrich}}, \bibinfo {author} {\bibfnamefont {L.}~\bibnamefont {Braun}},
  \bibinfo {author} {\bibfnamefont {J.~I.}\ \bibnamefont {Pascual}}, \ and\
  \bibinfo {author} {\bibfnamefont {K.~J.}\ \bibnamefont {Franke}},\
  }\href@noop {} {\bibfield  {journal} {\bibinfo  {journal} {Nature Physics}\
  }\textbf {\bibinfo {volume} {9}},\ \bibinfo {pages} {765} (\bibinfo {year}
  {2013})}\BibitemShut {NoStop}%
\bibitem [{\citenamefont {Nadj-Perge}\ \emph {et~al.}(2013)\citenamefont
  {Nadj-Perge}, \citenamefont {Drozdov}, \citenamefont {Bernevig},\ and\
  \citenamefont {Yazdani}}]{Nadj-Perdge13_Majorana_fermions_in_Shiba_chains}%
  \BibitemOpen
  \bibfield  {author} {\bibinfo {author} {\bibfnamefont {S.}~\bibnamefont
  {Nadj-Perge}}, \bibinfo {author} {\bibfnamefont {I.~K.}\ \bibnamefont
  {Drozdov}}, \bibinfo {author} {\bibfnamefont {B.~A.}\ \bibnamefont
  {Bernevig}}, \ and\ \bibinfo {author} {\bibfnamefont {A.}~\bibnamefont
  {Yazdani}},\ }\href {\doibase 10.1103/PhysRevB.88.020407} {\bibfield
  {journal} {\bibinfo  {journal} {Phys. Rev. B}\ }\textbf {\bibinfo {volume}
  {88}},\ \bibinfo {pages} {020407} (\bibinfo {year} {2013})}\BibitemShut
  {NoStop}%
\bibitem [{\citenamefont {Klinovaja}\ \emph {et~al.}(2013)\citenamefont
  {Klinovaja}, \citenamefont {Stano}, \citenamefont {Yazdani},\ and\
  \citenamefont
  {Loss}}]{Klinovaja13_TSC_and_Majorana_Fermions_in_RKKY_Systems}%
  \BibitemOpen
  \bibfield  {author} {\bibinfo {author} {\bibfnamefont {J.}~\bibnamefont
  {Klinovaja}}, \bibinfo {author} {\bibfnamefont {P.}~\bibnamefont {Stano}},
  \bibinfo {author} {\bibfnamefont {A.}~\bibnamefont {Yazdani}}, \ and\
  \bibinfo {author} {\bibfnamefont {D.}~\bibnamefont {Loss}},\ }\href {\doibase
  10.1103/PhysRevLett.111.186805} {\bibfield  {journal} {\bibinfo  {journal}
  {Phys. Rev. Lett.}\ }\textbf {\bibinfo {volume} {111}},\ \bibinfo {pages}
  {186805} (\bibinfo {year} {2013})}\BibitemShut {NoStop}%
\bibitem [{\citenamefont {Braunecker}\ and\ \citenamefont
  {Simon}(2013)}]{Braunecker13_Shiba_chain}%
  \BibitemOpen
  \bibfield  {author} {\bibinfo {author} {\bibfnamefont {B.}~\bibnamefont
  {Braunecker}}\ and\ \bibinfo {author} {\bibfnamefont {P.}~\bibnamefont
  {Simon}},\ }\href@noop {} {\bibfield  {journal} {\bibinfo  {journal} {Phys.
  Rev. Lett.}\ }\textbf {\bibinfo {volume} {111}},\ \bibinfo {pages} {147202}
  (\bibinfo {year} {2013})}\BibitemShut {NoStop}%
\bibitem [{\citenamefont {Pientka}\ \emph {et~al.}(2013)\citenamefont
  {Pientka}, \citenamefont {Glazman},\ and\ \citenamefont {von
  Oppen}}]{Pientka13_Shiba_chain}%
  \BibitemOpen
  \bibfield  {author} {\bibinfo {author} {\bibfnamefont {F.}~\bibnamefont
  {Pientka}}, \bibinfo {author} {\bibfnamefont {L.~I.}\ \bibnamefont
  {Glazman}}, \ and\ \bibinfo {author} {\bibfnamefont {F.}~\bibnamefont {von
  Oppen}},\ }\href@noop {} {\bibfield  {journal} {\bibinfo  {journal} {Phys.
  Rev. B}\ }\textbf {\bibinfo {volume} {88}},\ \bibinfo {pages} {155420}
  (\bibinfo {year} {2013})}\BibitemShut {NoStop}%
\bibitem [{\citenamefont {Nadj-Perge}\ \emph {et~al.}(2014)\citenamefont
  {Nadj-Perge}, \citenamefont {Drozdov}, \citenamefont {Li}, \citenamefont
  {Chen}, \citenamefont {Jeon}, \citenamefont {Seo}, \citenamefont {MacDonald},
  \citenamefont {Bernevig},\ and\ \citenamefont
  {Yazdani}}]{NadjPerge14_Observation_of_Majorana_fermions_in_Fe_chains}%
  \BibitemOpen
  \bibfield  {author} {\bibinfo {author} {\bibfnamefont {S.}~\bibnamefont
  {Nadj-Perge}}, \bibinfo {author} {\bibfnamefont {I.~K.}\ \bibnamefont
  {Drozdov}}, \bibinfo {author} {\bibfnamefont {J.}~\bibnamefont {Li}},
  \bibinfo {author} {\bibfnamefont {H.}~\bibnamefont {Chen}}, \bibinfo {author}
  {\bibfnamefont {S.}~\bibnamefont {Jeon}}, \bibinfo {author} {\bibfnamefont
  {J.}~\bibnamefont {Seo}}, \bibinfo {author} {\bibfnamefont {A.~H.}\
  \bibnamefont {MacDonald}}, \bibinfo {author} {\bibfnamefont {B.~A.}\
  \bibnamefont {Bernevig}}, \ and\ \bibinfo {author} {\bibfnamefont
  {A.}~\bibnamefont {Yazdani}},\ }\href@noop {} {\bibfield  {journal} {\bibinfo
   {journal} {Science}\ }\textbf {\bibinfo {volume} {346}},\ \bibinfo {pages}
  {602} (\bibinfo {year} {2014})}\BibitemShut {NoStop}%
\bibitem [{\citenamefont {Pawlak}\ \emph {et~al.}(2016)\citenamefont {Pawlak},
  \citenamefont {Kisiel}, \citenamefont {Klinovaja}, \citenamefont {Meier},
  \citenamefont {Kawai}, \citenamefont {Glatzel}, \citenamefont {Loss},\ and\
  \citenamefont
  {Meyer}}]{Pawlak15_Probing_Majorana_wavefunctions_in_Fe_chains}%
  \BibitemOpen
  \bibfield  {author} {\bibinfo {author} {\bibfnamefont {R.}~\bibnamefont
  {Pawlak}}, \bibinfo {author} {\bibfnamefont {M.}~\bibnamefont {Kisiel}},
  \bibinfo {author} {\bibfnamefont {J.}~\bibnamefont {Klinovaja}}, \bibinfo
  {author} {\bibfnamefont {T.}~\bibnamefont {Meier}}, \bibinfo {author}
  {\bibfnamefont {S.}~\bibnamefont {Kawai}}, \bibinfo {author} {\bibfnamefont
  {T.}~\bibnamefont {Glatzel}}, \bibinfo {author} {\bibfnamefont
  {D.}~\bibnamefont {Loss}}, \ and\ \bibinfo {author} {\bibfnamefont
  {E.}~\bibnamefont {Meyer}},\ }\href {http://dx.doi.org/10.1038/npjqi.2016.35}
  {\bibfield  {journal} {\bibinfo  {journal} {Npj Quantum Information}\
  }\textbf {\bibinfo {volume} {2}},\ \bibinfo {pages} {16035} (\bibinfo {year}
  {2016})}\BibitemShut {NoStop}%
\bibitem [{\citenamefont {Ruby}\ \emph
  {et~al.}(2015{\natexlab{b}})\citenamefont {Ruby}, \citenamefont {Pientka},
  \citenamefont {Peng}, \citenamefont {von Oppen}, \citenamefont {Heinrich},\
  and\ \citenamefont {Franke}}]{Ruby_2015}%
  \BibitemOpen
  \bibfield  {author} {\bibinfo {author} {\bibfnamefont {M.}~\bibnamefont
  {Ruby}}, \bibinfo {author} {\bibfnamefont {F.}~\bibnamefont {Pientka}},
  \bibinfo {author} {\bibfnamefont {Y.}~\bibnamefont {Peng}}, \bibinfo {author}
  {\bibfnamefont {F.}~\bibnamefont {von Oppen}}, \bibinfo {author}
  {\bibfnamefont {B.~W.}\ \bibnamefont {Heinrich}}, \ and\ \bibinfo {author}
  {\bibfnamefont {K.~J.}\ \bibnamefont {Franke}},\ }\href {\doibase
  10.1103/PhysRevLett.115.197204} {\bibfield  {journal} {\bibinfo  {journal}
  {Phys. Rev. Lett.}\ }\textbf {\bibinfo {volume} {115}},\ \bibinfo {pages}
  {197204} (\bibinfo {year} {2015}{\natexlab{b}})}\BibitemShut {NoStop}%
\bibitem [{\citenamefont {Kondo}(1964)}]{Kondo}%
  \BibitemOpen
  \bibfield  {author} {\bibinfo {author} {\bibfnamefont {J.}~\bibnamefont
  {Kondo}},\ }\href {\doibase 10.1143/PTP.32.37} {\bibfield  {journal}
  {\bibinfo  {journal} {Progress of Theoretical Physics}\ }\textbf {\bibinfo
  {volume} {32}},\ \bibinfo {pages} {37} (\bibinfo {year} {1964})}\BibitemShut
  {NoStop}%
\bibitem [{\citenamefont {Hewson}(1993)}]{Hewson_1993}%
  \BibitemOpen
  \bibfield  {author} {\bibinfo {author} {\bibfnamefont {A.~C.}\ \bibnamefont
  {Hewson}},\ }\href@noop {} {\emph {\bibinfo {title} {The Kondo Problem to
  Heavy Fermions}}}\ (\bibinfo  {publisher} {Cambridge University Press, New
  York},\ \bibinfo {year} {1993})\BibitemShut {NoStop}%
\bibitem [{\citenamefont {Zittartz}\ and\ \citenamefont
  {M{\"u}ller-Hartmann}(1970)}]{Zittartz_1970_I}%
  \BibitemOpen
  \bibfield  {author} {\bibinfo {author} {\bibfnamefont {J.}~\bibnamefont
  {Zittartz}}\ and\ \bibinfo {author} {\bibfnamefont {E.}~\bibnamefont
  {M{\"u}ller-Hartmann}},\ }\href {\doibase 10.1007/BF01394943} {\bibfield
  {journal} {\bibinfo  {journal} {Zeitschrift f{\"u}r Physik A Hadrons and
  nuclei}\ }\textbf {\bibinfo {volume} {232}},\ \bibinfo {pages} {11} (\bibinfo
  {year} {1970})}\BibitemShut {NoStop}%
\bibitem [{\citenamefont {Zittartz}(1970)}]{Zittartz_1970_III}%
  \BibitemOpen
  \bibfield  {author} {\bibinfo {author} {\bibfnamefont {J.}~\bibnamefont
  {Zittartz}},\ }\href@noop {} {\bibfield  {journal} {\bibinfo  {journal}
  {Zeitschrift f{\"u}r Physik A Hadrons and nuclei}\ }\textbf {\bibinfo
  {volume} {237}},\ \bibinfo {pages} {419} (\bibinfo {year}
  {1970})}\BibitemShut {NoStop}%
\bibitem [{\citenamefont {Yoshioka}\ and\ \citenamefont
  {Ohashi}(1998)}]{Yoshioka98_Kondo_impurity_in_SC_with_NRG}%
  \BibitemOpen
  \bibfield  {author} {\bibinfo {author} {\bibfnamefont {T.}~\bibnamefont
  {Yoshioka}}\ and\ \bibinfo {author} {\bibfnamefont {Y.}~\bibnamefont
  {Ohashi}},\ }\href@noop {} {\bibfield  {journal} {\bibinfo  {journal}
  {Journal Phys. Soc. Japan}\ }\textbf {\bibinfo {volume} {67}},\ \bibinfo
  {pages} {1332} (\bibinfo {year} {1998})}\BibitemShut {NoStop}%
\bibitem [{\citenamefont {Yoshioka}\ and\ \citenamefont
  {Ohashi}(2000)}]{Yoshioka00_NRG_Anderson_impurity_on_SC}%
  \BibitemOpen
  \bibfield  {author} {\bibinfo {author} {\bibfnamefont {T.}~\bibnamefont
  {Yoshioka}}\ and\ \bibinfo {author} {\bibfnamefont {Y.}~\bibnamefont
  {Ohashi}},\ }\href {\doibase 10.1143/JPSJ.69.1812} {\bibfield  {journal}
  {\bibinfo  {journal} {Journal of the Physical Society of Japan}\ }\textbf
  {\bibinfo {volume} {69}},\ \bibinfo {pages} {1812} (\bibinfo {year}
  {2000})}\BibitemShut {NoStop}%
\bibitem [{\citenamefont {Satori}\ \emph {et~al.}(1992)\citenamefont {Satori},
  \citenamefont {Shiba}, \citenamefont {Sakai},\ and\ \citenamefont
  {Shimizu}}]{Satori92_Magnetic_impurities_in_SC_with_NRG}%
  \BibitemOpen
  \bibfield  {author} {\bibinfo {author} {\bibfnamefont {K.}~\bibnamefont
  {Satori}}, \bibinfo {author} {\bibfnamefont {H.}~\bibnamefont {Shiba}},
  \bibinfo {author} {\bibfnamefont {O.}~\bibnamefont {Sakai}}, \ and\ \bibinfo
  {author} {\bibfnamefont {Y.}~\bibnamefont {Shimizu}},\ }\href {\doibase
  10.1143/JPSJ.61.3239} {\bibfield  {journal} {\bibinfo  {journal} {Journal of
  the Physical Society of Japan}\ }\textbf {\bibinfo {volume} {61}},\ \bibinfo
  {pages} {3239} (\bibinfo {year} {1992})}\BibitemShut {NoStop}%
\bibitem [{\citenamefont {Sakai}\ \emph {et~al.}(1993)\citenamefont {Sakai},
  \citenamefont {Shimizu}, \citenamefont {Shiba},\ and\ \citenamefont
  {Satori}}]{Sakai93_Magnetic_impurities_in_SC_with_NRG}%
  \BibitemOpen
  \bibfield  {author} {\bibinfo {author} {\bibfnamefont {O.}~\bibnamefont
  {Sakai}}, \bibinfo {author} {\bibfnamefont {Y.}~\bibnamefont {Shimizu}},
  \bibinfo {author} {\bibfnamefont {H.}~\bibnamefont {Shiba}}, \ and\ \bibinfo
  {author} {\bibfnamefont {K.}~\bibnamefont {Satori}},\ }\href {\doibase
  10.1143/JPSJ.62.3181} {\bibfield  {journal} {\bibinfo  {journal} {Journal of
  the Physical Society of Japan}\ }\textbf {\bibinfo {volume} {62}},\ \bibinfo
  {pages} {3181} (\bibinfo {year} {1993})}\BibitemShut {NoStop}%
\bibitem [{\citenamefont {Martín-Rodero}\ and\ \citenamefont
  {Levy~Yeyati}(2011)}]{MartinRodero11_Review_Josephson_and_Andreev_transport_through_QDs}%
  \BibitemOpen
  \bibfield  {author} {\bibinfo {author} {\bibfnamefont {A.}~\bibnamefont
  {Martín-Rodero}}\ and\ \bibinfo {author} {\bibfnamefont {A.}~\bibnamefont
  {Levy~Yeyati}},\ }\href@noop {} {\bibfield  {journal} {\bibinfo  {journal}
  {Advances in Physics}\ }\textbf {\bibinfo {volume} {60}},\ \bibinfo {pages}
  {899} (\bibinfo {year} {2011})}\BibitemShut {NoStop}%
\bibitem [{\citenamefont {\ifmmode~\check{Z}\else \v{Z}\fi{}itko}\ \emph
  {et~al.}(2011)\citenamefont {\ifmmode~\check{Z}\else \v{Z}\fi{}itko},
  \citenamefont {Bodensiek},\ and\ \citenamefont {Pruschke}}]{Zitko_2011}%
  \BibitemOpen
  \bibfield  {author} {\bibinfo {author} {\bibfnamefont {R.}~\bibnamefont
  {\ifmmode~\check{Z}\else \v{Z}\fi{}itko}}, \bibinfo {author} {\bibfnamefont
  {O.}~\bibnamefont {Bodensiek}}, \ and\ \bibinfo {author} {\bibfnamefont
  {T.}~\bibnamefont {Pruschke}},\ }\href {\doibase 10.1103/PhysRevB.83.054512}
  {\bibfield  {journal} {\bibinfo  {journal} {Phys. Rev. B}\ }\textbf {\bibinfo
  {volume} {83}},\ \bibinfo {pages} {054512} (\bibinfo {year}
  {2011})}\BibitemShut {NoStop}%
\bibitem [{\citenamefont {\ifmmode~\check{Z}\else
  \v{Z}\fi{}itko}(2018)}]{Zitko_2017}%
  \BibitemOpen
  \bibfield  {author} {\bibinfo {author} {\bibfnamefont {R.}~\bibnamefont
  {\ifmmode~\check{Z}\else \v{Z}\fi{}itko}},\ }\href {\doibase
  10.1016/j.physb.2017.08.019} {\bibfield  {journal} {\bibinfo  {journal}
  {Physica B: Condensed Matter}\ }\textbf {\bibinfo {volume} {536}},\ \bibinfo
  {pages} {230} (\bibinfo {year} {2018})}\BibitemShut {NoStop}%
\bibitem [{\citenamefont {Bauer}\ \emph {et~al.}(2007)\citenamefont {Bauer},
  \citenamefont {Oguri},\ and\ \citenamefont
  {Hewson}}]{Bauer07_NRG_Anderson_model_in_BCS_superconductor}%
  \BibitemOpen
  \bibfield  {author} {\bibinfo {author} {\bibfnamefont {J.}~\bibnamefont
  {Bauer}}, \bibinfo {author} {\bibfnamefont {A.}~\bibnamefont {Oguri}}, \ and\
  \bibinfo {author} {\bibfnamefont {A.~C.}\ \bibnamefont {Hewson}},\ }\href
  {http://stacks.iop.org/0953-8984/19/i=48/a=486211} {\bibfield  {journal}
  {\bibinfo  {journal} {Journal of Physics: Condensed Matter}\ }\textbf
  {\bibinfo {volume} {19}},\ \bibinfo {pages} {486211} (\bibinfo {year}
  {2007})}\BibitemShut {NoStop}%
\bibitem [{\citenamefont {\ifmmode~\check{Z}\else
  \v{Z}\fi{}itko}(2016)}]{Zitko16_Spectral_properties_of_Shiba_states_at_finite_T}%
  \BibitemOpen
  \bibfield  {author} {\bibinfo {author} {\bibfnamefont {R.}~\bibnamefont
  {\ifmmode~\check{Z}\else \v{Z}\fi{}itko}},\ }\href {\doibase
  10.1103/PhysRevB.93.195125} {\bibfield  {journal} {\bibinfo  {journal} {Phys.
  Rev. B}\ }\textbf {\bibinfo {volume} {93}},\ \bibinfo {pages} {195125}
  (\bibinfo {year} {2016})}\BibitemShut {NoStop}%
\bibitem [{\citenamefont {\ifmmode~\check{Z}\else \v{Z}\fi{}onda}\ \emph
  {et~al.}(2016)\citenamefont {\ifmmode~\check{Z}\else \v{Z}\fi{}onda},
  \citenamefont {Pokorn\'y}, \citenamefont {Jani\ifmmode~\check{s}\else
  \v{s}\fi{}},\ and\ \citenamefont
  {Novotn\'y}}]{Zonda16_Perturbation_theory_for_Anderson_impurity_in_SC}%
  \BibitemOpen
  \bibfield  {author} {\bibinfo {author} {\bibfnamefont {M.}~\bibnamefont
  {\ifmmode~\check{Z}\else \v{Z}\fi{}onda}}, \bibinfo {author} {\bibfnamefont
  {V.}~\bibnamefont {Pokorn\'y}}, \bibinfo {author} {\bibfnamefont
  {V.}~\bibnamefont {Jani\ifmmode~\check{s}\else \v{s}\fi{}}}, \ and\ \bibinfo
  {author} {\bibfnamefont {T.}~\bibnamefont {Novotn\'y}},\ }\href {\doibase
  10.1103/PhysRevB.93.024523} {\bibfield  {journal} {\bibinfo  {journal} {Phys.
  Rev. B}\ }\textbf {\bibinfo {volume} {93}},\ \bibinfo {pages} {024523}
  (\bibinfo {year} {2016})}\BibitemShut {NoStop}%
\bibitem [{\citenamefont {Jani{\v s}}\ \emph {et~al.}(2016)\citenamefont
  {Jani{\v s}}, \citenamefont {Pokorn{\'y}},\ and\ \citenamefont {{\v
  Z}onda}}]{Janis16_0_Pi_Transition_within_perturbation_theory}%
  \BibitemOpen
  \bibfield  {author} {\bibinfo {author} {\bibfnamefont {V.}~\bibnamefont
  {Jani{\v s}}}, \bibinfo {author} {\bibfnamefont {V.}~\bibnamefont
  {Pokorn{\'y}}}, \ and\ \bibinfo {author} {\bibfnamefont {M.}~\bibnamefont
  {{\v Z}onda}},\ }\href@noop {} {\bibfield  {journal} {\bibinfo  {journal}
  {Eur. Phys. J. B}\ }\textbf {\bibinfo {volume} {89}},\ \bibinfo {pages} {197}
  (\bibinfo {year} {2016})}\BibitemShut {NoStop}%
\bibitem [{\citenamefont {{\v Z}onda}\ \emph {et~al.}(2015)\citenamefont {{\v
  Z}onda}, \citenamefont {Pokorn{\'y}}, \citenamefont {Jani{\v s}},\ and\
  \citenamefont
  {Novotn{\'y}}}]{Zonda16_Perturbation_theory_of_SC_0_Pi_transition}%
  \BibitemOpen
  \bibfield  {author} {\bibinfo {author} {\bibfnamefont {M.}~\bibnamefont {{\v
  Z}onda}}, \bibinfo {author} {\bibfnamefont {V.}~\bibnamefont {Pokorn{\'y}}},
  \bibinfo {author} {\bibfnamefont {V.}~\bibnamefont {Jani{\v s}}}, \ and\
  \bibinfo {author} {\bibfnamefont {T.}~\bibnamefont {Novotn{\'y}}},\ }\href
  {\doibase doi:10.1038/srep08821} {\bibfield  {journal} {\bibinfo  {journal}
  {Scientific Reports}\ }\textbf {\bibinfo {volume} {5}},\ \bibinfo {pages}
  {8821} (\bibinfo {year} {2015})}\BibitemShut {NoStop}%
\bibitem [{\citenamefont {Luitz}\ and\ \citenamefont
  {Assaad}(2010)}]{Luitz10_QMC_study_of_Anderson_impurity_on_SC}%
  \BibitemOpen
  \bibfield  {author} {\bibinfo {author} {\bibfnamefont {D.~J.}\ \bibnamefont
  {Luitz}}\ and\ \bibinfo {author} {\bibfnamefont {F.~F.}\ \bibnamefont
  {Assaad}},\ }\href {\doibase 10.1103/PhysRevB.81.024509} {\bibfield
  {journal} {\bibinfo  {journal} {Phys. Rev. B}\ }\textbf {\bibinfo {volume}
  {81}},\ \bibinfo {pages} {024509} (\bibinfo {year} {2010})}\BibitemShut
  {NoStop}%
\bibitem [{\citenamefont
  {Tinkham}(1996)}]{Tinkham_Introduction_to_superconductivity}%
  \BibitemOpen
  \bibfield  {author} {\bibinfo {author} {\bibfnamefont {M.}~\bibnamefont
  {Tinkham}},\ }\href@noop {} {\emph {\bibinfo {title} {Introduction to
  Superconductivity, 2nd Edition}}}\ (\bibinfo  {publisher} {McGraw-Hill,
  Inc.},\ \bibinfo {address} {New York},\ \bibinfo {year} {1996})\BibitemShut
  {NoStop}%
\bibitem [{\citenamefont {Sakurai}(1970)}]{Sakurai70}%
  \BibitemOpen
  \bibfield  {author} {\bibinfo {author} {\bibfnamefont {A.}~\bibnamefont
  {Sakurai}},\ }\href {\doibase 10.1143/PTP.44.1472} {\bibfield  {journal}
  {\bibinfo  {journal} {Progress of Theoretical Physics}\ }\textbf {\bibinfo
  {volume} {44}},\ \bibinfo {pages} {1472} (\bibinfo {year}
  {1970})}\BibitemShut {NoStop}%
\bibitem [{\citenamefont {Soda}\ \emph {et~al.}(1967)\citenamefont {Soda},
  \citenamefont {Matsuura},\ and\ \citenamefont
  {Nagaoka}}]{Soda67_sd_exchange_in_a_SC}%
  \BibitemOpen
  \bibfield  {author} {\bibinfo {author} {\bibfnamefont {T.}~\bibnamefont
  {Soda}}, \bibinfo {author} {\bibfnamefont {T.}~\bibnamefont {Matsuura}}, \
  and\ \bibinfo {author} {\bibfnamefont {Y.}~\bibnamefont {Nagaoka}},\
  }\href@noop {} {\bibfield  {journal} {\bibinfo  {journal} {Progress of
  Theoretical Physics}\ }\textbf {\bibinfo {volume} {38}},\ \bibinfo {pages}
  {551} (\bibinfo {year} {1967})}\BibitemShut {NoStop}%
\bibitem [{\citenamefont {Nagaoka}(1965)}]{Nagaoka_1965}%
  \BibitemOpen
  \bibfield  {author} {\bibinfo {author} {\bibfnamefont {Y.}~\bibnamefont
  {Nagaoka}},\ }\href {\doibase 10.1103/PhysRev.138.A1112} {\bibfield
  {journal} {\bibinfo  {journal} {Phys. Rev.}\ }\textbf {\bibinfo {volume}
  {138}},\ \bibinfo {pages} {A1112} (\bibinfo {year} {1965})}\BibitemShut
  {NoStop}%
\bibitem [{\citenamefont {Fetter}\ and\ \citenamefont
  {Walecka}(1971)}]{fetter}%
  \BibitemOpen
  \bibfield  {author} {\bibinfo {author} {\bibfnamefont {A.~L.}\ \bibnamefont
  {Fetter}}\ and\ \bibinfo {author} {\bibfnamefont {J.~D.}\ \bibnamefont
  {Walecka}},\ }\href@noop {} {\emph {\bibinfo {title} {Quantum theory of
  many-particle systems}}}\ (\bibinfo  {publisher} {McGraw-Hill},\ \bibinfo
  {address} {New York},\ \bibinfo {year} {1971})\BibitemShut {NoStop}%
\bibitem [{\citenamefont {Fischer}\ \emph {et~al.}(2007)\citenamefont
  {Fischer}, \citenamefont {Kugler}, \citenamefont {Maggio-Aprile},
  \citenamefont {Berthod},\ and\ \citenamefont
  {Renner}}]{Fisher07_RMP_STM_HTSC}%
  \BibitemOpen
  \bibfield  {author} {\bibinfo {author} {\bibfnamefont {O.}~\bibnamefont
  {Fischer}}, \bibinfo {author} {\bibfnamefont {M.}~\bibnamefont {Kugler}},
  \bibinfo {author} {\bibfnamefont {I.}~\bibnamefont {Maggio-Aprile}}, \bibinfo
  {author} {\bibfnamefont {C.}~\bibnamefont {Berthod}}, \ and\ \bibinfo
  {author} {\bibfnamefont {C.}~\bibnamefont {Renner}},\ }\href {\doibase
  10.1103/RevModPhys.79.353} {\bibfield  {journal} {\bibinfo  {journal} {Rev.
  Mod. Phys.}\ }\textbf {\bibinfo {volume} {79}},\ \bibinfo {pages} {353}
  (\bibinfo {year} {2007})}\BibitemShut {NoStop}%
\bibitem [{\citenamefont {Zubarev}(1960)}]{Zubarev_1960}%
  \BibitemOpen
  \bibfield  {author} {\bibinfo {author} {\bibfnamefont {D.~N.}\ \bibnamefont
  {Zubarev}},\ }\href {http://stacks.iop.org/0038-5670/3/i=3/a=R02} {\bibfield
  {journal} {\bibinfo  {journal} {Soviet Physics Uspekhi}\ }\textbf {\bibinfo
  {volume} {3}},\ \bibinfo {pages} {320} (\bibinfo {year} {1960})}\BibitemShut
  {NoStop}%
\bibitem [{\citenamefont {Costi}(2000)}]{Costi00}%
  \BibitemOpen
  \bibfield  {author} {\bibinfo {author} {\bibfnamefont {T.~A.}\ \bibnamefont
  {Costi}},\ }\href {\doibase 10.1103/PhysRevLett.85.1504} {\bibfield
  {journal} {\bibinfo  {journal} {Phys. Rev. Lett.}\ }\textbf {\bibinfo
  {volume} {85}},\ \bibinfo {pages} {1504} (\bibinfo {year}
  {2000})}\BibitemShut {NoStop}%
\bibitem [{\citenamefont {M{\'e}nard}\ \emph {et~al.}(2015)\citenamefont
  {M{\'e}nard}, \citenamefont {Guissart}, \citenamefont {Brun}, \citenamefont
  {Pons}, \citenamefont {Stolyarov}, \citenamefont {Debontridder},
  \citenamefont {Leclerc}, \citenamefont {Janod}, \citenamefont {Cario},
  \citenamefont {Roditchev}, \citenamefont {Simon},\ and\ \citenamefont
  {Cren}}]{Menard2015}%
  \BibitemOpen
  \bibfield  {author} {\bibinfo {author} {\bibfnamefont {G.~C.}\ \bibnamefont
  {M{\'e}nard}}, \bibinfo {author} {\bibfnamefont {S.}~\bibnamefont
  {Guissart}}, \bibinfo {author} {\bibfnamefont {C.}~\bibnamefont {Brun}},
  \bibinfo {author} {\bibfnamefont {S.}~\bibnamefont {Pons}}, \bibinfo {author}
  {\bibfnamefont {V.~S.}\ \bibnamefont {Stolyarov}}, \bibinfo {author}
  {\bibfnamefont {F.}~\bibnamefont {Debontridder}}, \bibinfo {author}
  {\bibfnamefont {M.~V.}\ \bibnamefont {Leclerc}}, \bibinfo {author}
  {\bibfnamefont {E.}~\bibnamefont {Janod}}, \bibinfo {author} {\bibfnamefont
  {L.}~\bibnamefont {Cario}}, \bibinfo {author} {\bibfnamefont
  {D.}~\bibnamefont {Roditchev}}, \bibinfo {author} {\bibfnamefont
  {P.}~\bibnamefont {Simon}}, \ and\ \bibinfo {author} {\bibfnamefont
  {T.}~\bibnamefont {Cren}},\ }\href {https://doi.org/10.1038/nphys3508}
  {\bibfield  {journal} {\bibinfo  {journal} {Nature Physics}\ }\textbf
  {\bibinfo {volume} {11}},\ \bibinfo {pages} {1013} (\bibinfo {year}
  {2015})}\BibitemShut {NoStop}%
\bibitem [{\citenamefont {Kezilebieke}\ \emph {et~al.}(2018)\citenamefont
  {Kezilebieke}, \citenamefont {Dvorak}, \citenamefont {Ojanen},\ and\
  \citenamefont {Liljeroth}}]{Kezilebieke2018}%
  \BibitemOpen
  \bibfield  {author} {\bibinfo {author} {\bibfnamefont {S.}~\bibnamefont
  {Kezilebieke}}, \bibinfo {author} {\bibfnamefont {M.}~\bibnamefont {Dvorak}},
  \bibinfo {author} {\bibfnamefont {T.}~\bibnamefont {Ojanen}}, \ and\ \bibinfo
  {author} {\bibfnamefont {P.}~\bibnamefont {Liljeroth}},\ }\href {\doibase
  10.1021/acs.nanolett.7b05050} {\bibfield  {journal} {\bibinfo  {journal}
  {Nano Letters}\ }\textbf {\bibinfo {volume} {18}},\ \bibinfo {pages} {2311}
  (\bibinfo {year} {2018})}\BibitemShut {NoStop}%
\bibitem [{\citenamefont {Heinrich}\ \emph {et~al.}(2015)\citenamefont
  {Heinrich}, \citenamefont {Braun}, \citenamefont {Pascual},\ and\
  \citenamefont {Franke}}]{Heinrich_2015}%
  \BibitemOpen
  \bibfield  {author} {\bibinfo {author} {\bibfnamefont {B.~W.}\ \bibnamefont
  {Heinrich}}, \bibinfo {author} {\bibfnamefont {L.}~\bibnamefont {Braun}},
  \bibinfo {author} {\bibfnamefont {J.~I.}\ \bibnamefont {Pascual}}, \ and\
  \bibinfo {author} {\bibfnamefont {K.~J.}\ \bibnamefont {Franke}},\ }\href
  {\doibase 10.1021/acs.nanolett.5b00987} {\bibfield  {journal} {\bibinfo
  {journal} {Nano Letters}\ }\textbf {\bibinfo {volume} {15}},\ \bibinfo
  {pages} {4024} (\bibinfo {year} {2015})},\ \bibinfo {note} {pMID: 25942560},\
  \Eprint {http://arxiv.org/abs/http://dx.doi.org/10.1021/acs.nanolett.5b00987}
  {http://dx.doi.org/10.1021/acs.nanolett.5b00987} \BibitemShut {NoStop}%
\bibitem [{\citenamefont {Kir\ifmmode~\check{s}\else \v{s}\fi{}anskas}\ \emph
  {et~al.}(2015)\citenamefont {Kir\ifmmode~\check{s}\else \v{s}\fi{}anskas},
  \citenamefont {Goldstein}, \citenamefont {Flensberg}, \citenamefont
  {Glazman},\ and\ \citenamefont
  {Paaske}}]{Kirsanskas15_YSR_states_in_phase_biased_SC_QD_SC_junctions}%
  \BibitemOpen
  \bibfield  {author} {\bibinfo {author} {\bibfnamefont {G.}~\bibnamefont
  {Kir\ifmmode~\check{s}\else \v{s}\fi{}anskas}}, \bibinfo {author}
  {\bibfnamefont {M.}~\bibnamefont {Goldstein}}, \bibinfo {author}
  {\bibfnamefont {K.}~\bibnamefont {Flensberg}}, \bibinfo {author}
  {\bibfnamefont {L.~I.}\ \bibnamefont {Glazman}}, \ and\ \bibinfo {author}
  {\bibfnamefont {J.}~\bibnamefont {Paaske}},\ }\href {\doibase
  10.1103/PhysRevB.92.235422} {\bibfield  {journal} {\bibinfo  {journal} {Phys.
  Rev. B}\ }\textbf {\bibinfo {volume} {92}},\ \bibinfo {pages} {235422}
  (\bibinfo {year} {2015})}\BibitemShut {NoStop}%
\bibitem [{\citenamefont {Hirjibehedin}\ \emph {et~al.}(2007)\citenamefont
  {Hirjibehedin}, \citenamefont {Lin}, \citenamefont {Otte}, \citenamefont
  {Ternes}, \citenamefont {Lutz}, \citenamefont {Jones},\ and\ \citenamefont
  {Heinrich}}]{Hirjibehedin07_Magnetic_anisotropy_in_surfaces}%
  \BibitemOpen
  \bibfield  {author} {\bibinfo {author} {\bibfnamefont {C.~F.}\ \bibnamefont
  {Hirjibehedin}}, \bibinfo {author} {\bibfnamefont {C.-Y.}\ \bibnamefont
  {Lin}}, \bibinfo {author} {\bibfnamefont {A.~F.}\ \bibnamefont {Otte}},
  \bibinfo {author} {\bibfnamefont {M.}~\bibnamefont {Ternes}}, \bibinfo
  {author} {\bibfnamefont {C.~P.}\ \bibnamefont {Lutz}}, \bibinfo {author}
  {\bibfnamefont {B.~A.}\ \bibnamefont {Jones}}, \ and\ \bibinfo {author}
  {\bibfnamefont {A.~J.}\ \bibnamefont {Heinrich}},\ }\href@noop {} {\bibfield
  {journal} {\bibinfo  {journal} {Science}\ }\textbf {\bibinfo {volume}
  {317}},\ \bibinfo {pages} {1199} (\bibinfo {year} {2007})}\BibitemShut
  {NoStop}%
\bibitem [{\citenamefont {\v{Z}itko}\ and\ \citenamefont
  {Pruschke}(2010)}]{Zitko10_IETS_for_Fe_atoms}%
  \BibitemOpen
  \bibfield  {author} {\bibinfo {author} {\bibfnamefont {R.}~\bibnamefont
  {\v{Z}itko}}\ and\ \bibinfo {author} {\bibfnamefont {T.}~\bibnamefont
  {Pruschke}},\ }\href@noop {} {\bibfield  {journal} {\bibinfo  {journal} {New
  J. Phys.}\ }\textbf {\bibinfo {volume} {12}},\ \bibinfo {pages} {063040}
  (\bibinfo {year} {2010})}\BibitemShut {NoStop}%
\bibitem [{\citenamefont {Otte}\ \emph {et~al.}(2008)\citenamefont {Otte},
  \citenamefont {Ternes}, \citenamefont {von Bergmann}, \citenamefont {Loth},
  \citenamefont {Brune}, \citenamefont {Lutz}, \citenamefont {Hirjibehedin},\
  and\ \citenamefont
  {Heinrich}}]{Otte08_The_role_of_anisotropy_on_Kondo_effect}%
  \BibitemOpen
  \bibfield  {author} {\bibinfo {author} {\bibfnamefont {A.~F.}\ \bibnamefont
  {Otte}}, \bibinfo {author} {\bibfnamefont {M.}~\bibnamefont {Ternes}},
  \bibinfo {author} {\bibfnamefont {K.}~\bibnamefont {von Bergmann}}, \bibinfo
  {author} {\bibfnamefont {S.}~\bibnamefont {Loth}}, \bibinfo {author}
  {\bibfnamefont {H.}~\bibnamefont {Brune}}, \bibinfo {author} {\bibfnamefont
  {C.~P.}\ \bibnamefont {Lutz}}, \bibinfo {author} {\bibfnamefont {C.~F.}\
  \bibnamefont {Hirjibehedin}}, \ and\ \bibinfo {author} {\bibfnamefont
  {A.~J.}\ \bibnamefont {Heinrich}},\ }\href
  {https://doi.org/10.1038/nphys1072} {\bibfield  {journal} {\bibinfo
  {journal} {Nature Physics}\ }\textbf {\bibinfo {volume} {4}},\ \bibinfo
  {pages} {847} (\bibinfo {year} {2008})}\BibitemShut {NoStop}%
\bibitem [{\citenamefont {Abramowitz}\ and\ \citenamefont
  {Stegun}(1965)}]{abramowitz}%
  \BibitemOpen
  \bibfield  {author} {\bibinfo {author} {\bibfnamefont {M.}~\bibnamefont
  {Abramowitz}}\ and\ \bibinfo {author} {\bibfnamefont {I.~A.}\ \bibnamefont
  {Stegun}},\ }\href@noop {} {\emph {\bibinfo {title} {Handbook of mathematical
  functions : with formulas, graphs and mathematical tables}}}\ (\bibinfo
  {publisher} {Dover},\ \bibinfo {address} {New York},\ \bibinfo {year}
  {1965})\BibitemShut {NoStop}%
\end{thebibliography}
%

\end{document}